\documentclass[usenatbib,useAMS]{mn2e}

\usepackage{graphicx,times}

\newcommand{\Hb}{\ensuremath{{\rm H}\beta}}

\newcommand{\HgA}{\ensuremath{{\rm H}\gamma_{\rm A}}}
\newcommand{\HgF}{\ensuremath{{\rm H}\gamma_{\rm F}}}

\newcommand{\HdA}{\ensuremath{{\rm H}\delta_{\rm A}}}

\newcommand{\CNone}{\ensuremath{{\rm CN}_1}}
\newcommand{\CNtwo}{\ensuremath{{\rm CN}_2}}
\newcommand{\Ctwo}{\ensuremath{{\rm C}_2}4668}
\newcommand{\Mgone}{\ensuremath{{\rm Mg}_1}}
\newcommand{\Mgtwo}{\ensuremath{{\rm Mg}_2}}
\newcommand{\Mgb}{\ensuremath{{\rm Mg}\, b}}

\newcommand{\aFe}{\ensuremath{\alpha/{\rm Fe}}}
\newcommand{\CaFe}{\ensuremath{{\rm Ca}/{\rm Fe}}}
\newcommand{\TiFe}{\ensuremath{{\rm Ti}/{\rm Fe}}}

\newcommand{\FeH}{\ensuremath{{\rm Fe}/{\rm H}}}
\newcommand{\MgFep}{\ensuremath{[{\rm MgFe}]^{\prime}}}
\newcommand{\ZH}{\ensuremath{Z/{\rm H}}}
\newcommand{\MgFe}{\ensuremath{{\rm Mg}/{\rm Fe}}}
\newcommand{\OFe}{\ensuremath{{\rm O}/{\rm Fe}}}
\newcommand{\NFe}{\ensuremath{{\rm N}/{\rm Fe}}}
\newcommand{\CFe}{\ensuremath{{\rm C}/{\rm Fe}}}

\title[Chemical abundance ratios of galactic globular clusters]
{Chemical abundance ratios of galactic globular clusters from modelling integrated light spectroscopy}

\author[Thomas, Johansson, Maraston] {
\parbox[h]{\textwidth}{Daniel Thomas, Jonas Johansson, Claudia Maraston}
\vspace*{8pt}\\ 
Institute of Cosmology and Gravitation, University of Portsmouth, Dennis Sciama Building, Burnaby Road, Portsmouth, PO1 3FX, UK\\
SEPNET, South East Physics Network}

\date{Accepted ... Received 20 October 2010 ; in original form 14 July 2010}

\pagerange{\pageref{firstpage}--\pageref{lastpage}}

\pubyear{2010}

\begin{document}

\bibliographystyle{mn2e}
\maketitle

\label{firstpage}

\begin{abstract}
In a companion paper we present new, flux-calibrated stellar population models of Lick absorption-line indices with variable element abundance ratios. The model includes a large variety of individual element variations, which allows the derivation of  the abundances for the elements C, N, O, Mg, Ca, Ti, and Fe besides total metallicity and age. We use this model to obtain estimates of these quantities from integrated light spectroscopy of galactic globular clusters. We show that the model fits to a number of indices improve considerably when various variable element ratios are considered. The ages we derive agree well with the literature and are all consistent with the age of the universe within the measurement errors. There is a considerable scatter in the ages, though, and we overestimate the ages preferentially for the metal-rich globular clusters. Our derived total metallicities agree generally very well with literature values on the \citet{ZW84} scale once corrected for $\alpha$-enhancement, in particular for those cluster where the ages agree with the CMD ages. We tend to slightly underestimate the metallicity for those clusters where we overestimate the age, in line with the age-metallicity degeneracy.
It turns out that the derivation of individual element abundance ratios is not reliable at an iron abundance $[\FeH]<-1\;$dex where line strengths become weaker, while the [\aFe] ratio is robust at all metallicities. The discussion of individual element ratios focuses therefore on globular clusters with $[\FeH]>-1\;$dex. We find general enhancement of light and $\alpha$ elements, as expected, with significant variations for some elements. The elements O and Mg follow the same general enhancement with almost identical distributions of [O/Fe] and [Mg/Fe]. We obtain slightly lower [C/Fe] and very high [N/Fe] ratios, instead. This chemical anomaly, commonly attributed to self-enrichment, is well known in globular clusters from individual stellar spectroscopy. It is the first time that this pattern is obtained also from the integrated light. The $\alpha$ elements follow a pattern such that the heavier elements Ca and Ti are less enhanced. More specifically, the [Ca/Fe] and [Ti/Fe] ratios are lower than [O/Fe] and [Mg/Fe] by about $0.2\;$dex. Most interestingly this trend of element abundance with atomic number is also seen in recent determinations of element abundances in globular cluster and field stars of the Milky Way. This suggests that Type~Ia supernovae contribute significantly to the enrichment of the heavier $\alpha$ elements as predicted by nucleosynthesis calculations and galactic chemical evolution models.
\end{abstract}



\section{Introduction}
The abundances of a large variety of chemical elements can be derived from high-resolution spectroscopy of individual stars in the field and globular clusters of the Milky Way as well as nearby dwarf galaxies in the Local Group \citep[e.g.,][]{McWilliam97,Carretta05,Pritzl05,Tolstoy09,Bensby10}. This level of detail cannot be achieved for most galaxies and extra-galactic globular clusters, because the individual stars are not resolved. Observations have to resort to integrated light spectroscopy, which is applicable to unresolved stellar populations. It allows us to study element abundances in distant galaxies and globular clusters, but is naturally more limited. Nearby globular clusters are the interface between these two extremes. They allow detailed chemical analyses from resolved stellar spectroscopy as well as the study of their integrated light. They are therefore vital for the calibration of stellar population models and integrated light analyses.

The Lick group have defined a set of 25 optical absorption-line indices \citep{Burstein84,Faber85,Gorgas93,Wortheyetal94,Trager98,WortheyOttaviani97}, the so-called Lick index system, that are by far the most commonly used in absorption-line analyses of old stellar populations. These can be used, in principle, for the derivation of various element abundance ratios. The bandpasses of Lick indices are relatively large with widths up to $50\;$\AA, which increases the signal-to-noise ratios but complicates their use for the derivation of individual element abundances. The first stellar population models of Lick absorption-line indices with variable element abundance ratios have been published by \citet[][hereafter TMB/K models]{Thomas03a,TMK04} based on the index response functions by \citet{TB95} and \citet{KMT05}. In a companion paper \citep[][hereafter Paper~I]{Thomas10b} we have updated these models (hereafter TMJ models) that are now flux calibrated thanks to the use of the newly computed index calibrations by \citet{JTM10} based on the flux-calibrated stellar library MILES \citep{Sanchez06a}.

The \citet{KMT05} model atmosphere calculations provide index response functions for the variation of the ten elements C, N, O, Mg, Na, Si, Ca, Ti, Fe, and Cr. Through additional features in the same part of the spectrum and modifications of the index definitions even more elements may be accessible \citep{Serven05,Lee09a}. Here we focus on those elements that can be best derived from the 25 Lick indices considered in the TMJ models. These are C, N, Mg, Ca, Ti, and Fe besides total metallicity \ZH, age, and \aFe\ ratio.

We already used the \citet{Thomas03a} code to derive abundances of nitrogen and calcium for globular clusters and galaxies. Through a simple approach in \citet{Thomas03a} we could show that galactic globular clusters must be significantly enhanced in nitrogen at fixed carbon abundance in order to reproduce the observed CN indices. In \citet{Thomas03b} we derive calcium abundances of galaxies. Subsequent work has developed this further. \citet{Clemens06} add C abundances in their study of SDSS galaxies, and \citet{Kelson06} derive N abundances of distant galaxies. \citet{Graves08} and \citet{Smith09} present the first full analyses of the abundances of C, N, Mg, Ca, and Fe in galaxies. In this paper we conduct the next step by adding the element titanium and derive the element abundance ratios [C/Fe], [N/Fe], [O/Fe], [Mg/Fe], [Ca/Fe], and [Ti/Fe] for galactic globular clusters.

The paper is organised as follows. In Section~\ref{sec:data} we describe the globular cluster data used. In Section~\ref{sec:model} we introduce the new TMJ model and outline our method to derive element ratios. The main analysis is presented in Section~\ref{sec:results}. The results are discussed in Section~\ref{sec:discussion}, and the paper concludes with Section~\ref{sec:conclusions}.

\section{Globular cluster data}
\label{sec:data}
Following our strategy for the TMB/K models, in Paper~I we compare the model predictions with observational data of galactic globular clusters, as the latter are the closest analogues of simple stellar populations in the real universe \citep{Maraston03a}. Key is that independent estimates of ages, metallicities, and element abundance ratios are available for the globular clusters of the Milky Way from deep photometry and high-resolution stellar spectroscopy.

The globular cluster samples are from \citet[][hereafter P02]{Puzia02} and \citet[][hereafter S05]{Schiavon05}. Critical for the integrated light spectroscopy is a representative sampling of the underlying stellar population \citep{Renzini98,Maraston98}. To ensure this P02 obtained several spectra with slightly offset pointings. In general three long-slit spectra were taken for each of the target clusters, and the observing pattern was optimized to obtain one spectrum of the nuclear region and spectra of adjacent fields. Exposure times were adjusted according to the surface brightness of each globular cluster to reach a statistically secure luminosity sampling of the underlying stellar population. S05, instead, obtained each observation by drifting the spectrograph slit across the core diameter of the cluster. The telescope was positioned so as to offset the slit from the cluster center by one core radius. A suitable trail rate was chosen to allow the slit to drift across the cluster core diameter during the typically 15 minute long exposure.

We do not use the indices tabulated in P02 directly, because these measurements have been  calibrated onto the Lick/IDS system by correcting for Lick offsets. S05 do not provide line index measurements. Hence we measure line strengths of all 25 Lick absorption-line indices for both samples directly on the globular cluster spectra using the definitions by \citet{Trager98}. Both globular cluster samples have been flux calibrated, so that no further offsets need to be applied for the comparison with the TMJ models. We have smoothed the spectra to Lick spectral resolution before the index measurement. Note that the spectral resolutions of both samples are below the resolution of the MILES library, so that we work with the TMJ models at Lick resolution.

For the P02 sample we adopt the errors quoted in their paper using the quadratic sum of the statistical (Poisson) error, the statistical error derived from slit to slit variations, and the systematic error introduced through uncertainties in the radial velocity. This information is not directly available from S05. We therefore evaluate the measurement errors in two steps. First we compute the Poisson errors from the error spectra provided through Monte Carlo simulations. Then we scale these errors with the complete errors from P02 from the overlapping globular clusters.

We add the slit-to-slit error evaluated in P02 in order to account for possible stellar population fluctuations that are not included in the statistical error. The observing strategies in both P02 and S05 have been designed to minimise such an error. Still, this effect may not negligible. The slit-to-slit variations overestimate this effect, as P02 have typically observed three slits per cluster. We regard the errors used in this study therefore as conservative estimate, and true errors are likely to be smaller.

Finally, it should be noted that the S05 spectra are corrupt around 4546 and $5050\;$\AA, so that the indices Fe4531 and Fe5015 cannot be measured (S05). In case of multiple observations in S05 we use the spectra with the highest signal-to-noise ratio.

\section{The TMJ model}
\label{sec:model}
In Paper~I we present new stellar population models of Lick absorption-line indices with variable element abundance ratios (TMJ). The model is an extension of the TMB/K model, which is based on the evolutionary stellar population synthesis code of \citet{Maraston98,Maraston05}. For basic information on the model we refer the reader to \citet{Thomas03a,TMK04} and Paper~I. Here we provide a brief summary of the main features of our new models.

\subsection{New features}
The key novelty compared to our previous models is that the TMJ model is flux-calibrated, hence not tied anymore to the Lick/IDS system. This is because the new models are based on our calibrations of absorption-line indices with stellar parameters \citep{JTM10} derived from the flux-calibrated stellar library MILES \citep{Sanchez06a}. The MILES library consists of 985 stars selected to produce a sample with extensive stellar parameter coverage. Most importantly it has been carefully flux-calibrated, making standard star-derived offsets unnecessary.

A further new feature is that we provide model predictions for both the original Lick ($\sim 8\;$\AA) and the higher MILES ($\sim 2.7\;$\AA) spectral resolutions. Note that the latter appears to be comparable to the SDSS resolution, so that our new high-resolution models can be applied to SDSS data without any corrections for instrumental spectral resolution (see Paper~I).

As a further novelty we calculate statistical errors in the model predictions. 
The errors estimates are obtained from the uncertainties in the measurements of Lick index strengths and the stellar parameters of the library stars, hence do not include systematic errors.
It turns out that the model errors are generally very small and well below the observational errors around solar metallicity, but rise considerably toward the highest and lowest metallicities.

The data release now provides models with two different stellar evolutionary tracks by \citet{Cassisi97} as used in TMB/K and additionally Padova \citep{Girardi00} at high metallicities. The model based on the Padova tracks is consistent with the model using Cassisi for the majority of indices. The cases of indices for which the discrepancy exceeds the model error significantly are \Hb, \CNone, \CNtwo, and \Ctwo. Small deviations are found for Ca4227, G4300, Ca4455, Fe5015, and Fe5709. In all cases, equally for the Balmer and the metal lines, the Padova based models produce lower index strengths. This effects kicks in at super-solar metallicities, however, outside the range of globular cluster metallicities. In the present study we use the models based on the Cassisi tracks. Finally, most importantly for the present paper, we release additional model tables with enhancement of each of the elements C, N, Na, Mg, Si, Ca, Ti, and Cr separately by $0.3\;$dex. A differential element ratio bias at low metallicities is considered to account of the fact that heavier $\alpha$ elements like Ca and Ti tend to be less enhanced (see Section~\ref{sec:discussion}).

The basic tests with globular cluster data following our strategy for the TMB/K model is presented in Paper~I. The match to the globular cluster data is satisfactory for the Balmer line indices  \HdA, \HgA, and \HgF. A reasonably good match with globular cluster data is also seen for the \aFe\ sensitive, metallic indices G4300, \Mgtwo, and \Mgb\ and the Fe indices Fe4383, Fe4531, Fe5270, Fe5335, Fe5406. The models are well off, instead, for the indices \CNone, \CNtwo, Ca4227, \Ctwo, and \Mgone. This is caused by the variation of further chemical elements beyond the \aFe\ ratio to which these indices are sensitive, and the full analysis of these element abundance variations is subject of the present work. In the following section we describe how individual element ratios are derived from this set of absorption features.

\begin{figure*}
\includegraphics[angle=-90,width=0.7\textwidth]{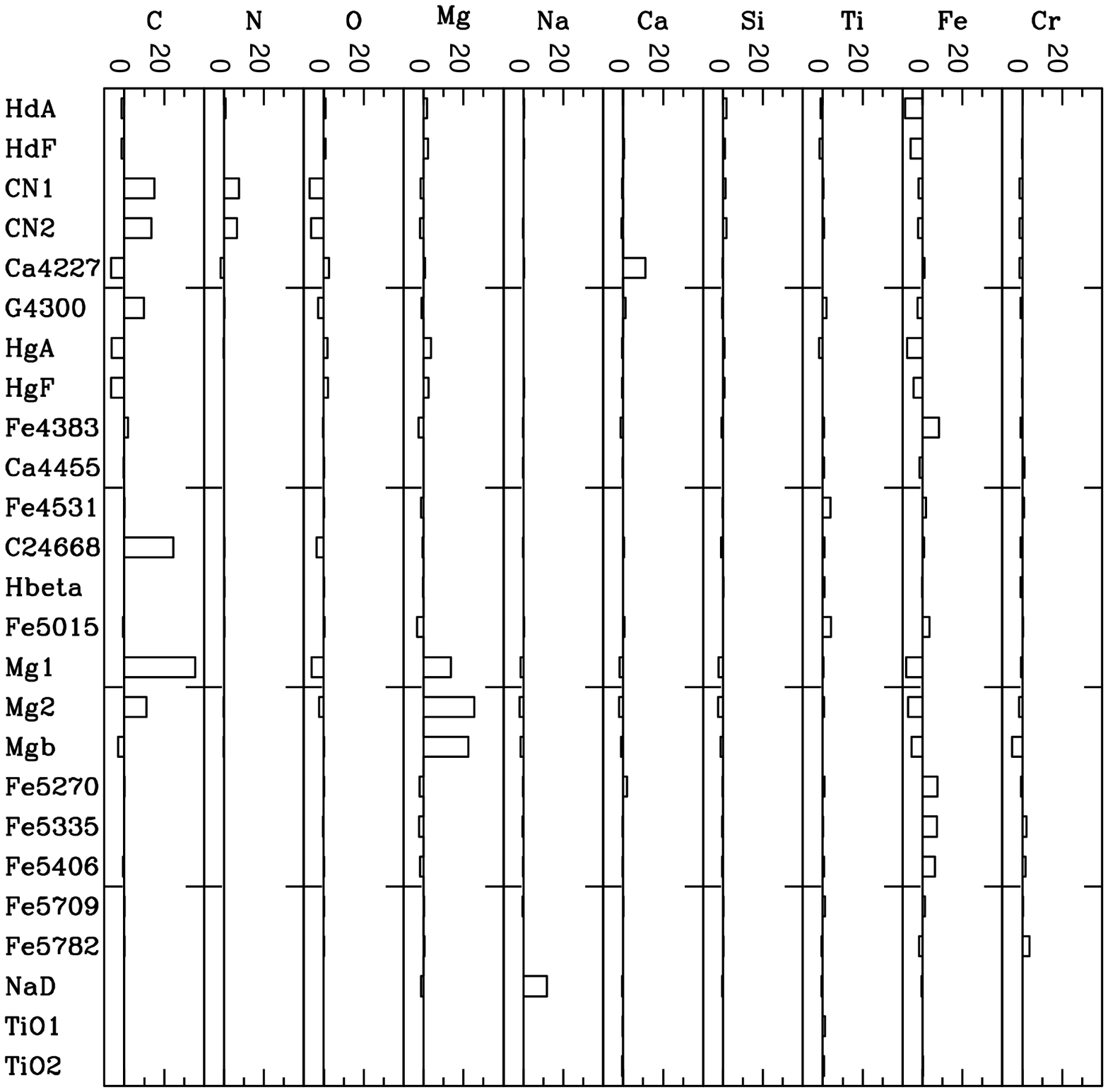}
\caption{Response of the 25 Lick indices to individual element abundance changes for a $12\;$Gyr, solar metallicity stellar population. The fractional index change is calculated for an enhancement of the respective element by a factor of two. The scale on the y-axis is error normalised. The plot range is kept fixed for all elements, so that the figure allows us to identify easily those elements that are best traced by the current set of models.}
\label{fig:response}
\end{figure*}

\subsection{Index responses}

Fig.~\ref{fig:response} shows the response of the 25 Lick indices to individual element abundance changes for a $12\;$Gyr, solar metallicity stellar population. The fractional index change is calculated for an enhancement of the respective element by a factor two normalised to the typical observational measurement error for MILES stars from \citet{JTM10}. The scale on the y-axis is kept fixed for all elements, so that the figure allows us to identify easily those elements that are best traced by the current set of models. It can be seen that the elements C, N, Na, Mg, Ca, Ti, and Fe are best accessible.

The abundance of nitrogen is obtained from the CN indices that are also highly sensitive to C abundance. However, this degeneracy can be easily broken through other C sensitive indices such as \Ctwo\ and \Mgone. The Mg indices \Mgone, \Mgtwo, and \Mgb\ are very sensitive to Mg abundance. Note, however, that all three additionally anti-correlate with Fe abundance \citep{Trager00a,Thomas03a}. Ca can be measured well from Ca4227, except that this particular index is quite weak and requires good data quality. Na abundance can be derived quite easily from NaD in principle. However, in practise this is problematic as the stellar component of this absorption feature is highly contaminated by interstellar absorption, which makes this index useless and hence Na inaccessible at least for globular clusters \citep{Thomas03a}. Iron is well sampled through the Fe indices.

There are two among the Fe indices, however, that are also sensitive to Ti abundance besides Fe. These are Fe4531 and Fe5015. They offer the opportunity to estimate also Ti abundance. We will only use Fe4531, as Fe5015 is contaminated by a non-negligible Mg sensitivity besides Fe, which weakens its usefulness for Ti abundance determinations.

The remaining three elements O, Si and Cr cannot easily be measured through the available indices. As discussed extensively in \citet{Thomas03a}, however, oxygen has a special role. O is by far the most abundant metal and clearly dominates the mass budget of 'total metallicity'. Moreover, the \aFe\ ratio is actually characterised by a depression in Fe abundance relative to all light elements (not only the $\alpha$ elements), hence \aFe\ reflects the ratio between total metallicity to iron ratio rather than $\alpha$ element abundance to iron. As total metallicity is driven by oxygen abundance, the \aFe\ derived can be most adequately interpreted as O/Fe ratio. We therefore re-name the parameter \aFe\ to O/Fe under  the assumption that this ratio provides an indirect measurement of oxygen abundance.

Finally, iron abundance can be calculated through the following formula \citep{Tantalo98,Trager00a,Thomas03a}.

\[ [{\rm Fe}/{\rm H}]=[Z/H]-0.93\: [\aFe]\equiv [Z/H]-0.93\: [\OFe]  \]

\section{Method}
A full description of the method we use to derive individual element abundances and its application to SDSS galaxy data is presented in a companion paper (Johansson et al, in preparation). Here we provide a brief summary of the key aspects that are most relevant for the present work.

\subsection{\boldmath The $\chi^{2}$ technique}
The derivation of the above set of element ratios is done in various iterative steps by means of the $\chi^{2}$ code of \citet{Thomas10a}. We generate a fine grid of model predictions for the parameters log age, metallicity, and \aFe\ ratio with log steps of 0.1, 0.1, and $0.05\;$dex, respectively. Galactic globular cluster data generally are very close to the 15$\;$Gyr model (see Paper~I), which is the highest age for which we have stellar evolutionary track calculations available. Therefore, we extrapolate the models logarithmically to a maximum age of $20\;$Gyr for the initial set of templates, in order not to impose an upper age limit. Note that the index strengths evolve very little as a function of age at these old ages, therefore we do not expect this extrapolation to affect the derivation of individual element abundances significantly. As a sanity check we have verified that the globular clusters with ages above $15\;$Gyr are not biased to particular element abundance ratios.

The code computes the $\chi^{2}$ between model prediction and observed index value for all model templates summing over the $n$ indices considered:

\begin{equation}
\chi^{2}=\sum_{i=1}^{n} \left(\frac{I_{i}^{\rm obs}-I_{i}^{\rm model}}{\sigma}\right)^{2}
\end{equation}
The resulting $\chi^{2}$ distribution is then transformed into a probability distribution. By means of the incomplete $\Gamma$ function adopting the degrees of freedom as $\nu=n_{\rm indices}-n_{\rm para}$ we compute the probability $Q$ that the chi-square should exceed a particular value $\chi^{2}$ by chance. This computed probability gives a quantitative measure for the goodness-of-fit of the model. If $Q$ is a very small probability, then the apparent discrepancies are unlikely to be chance fluctuations.

The solution with the highest $Q$ (i.e.\ lowest $\chi^{2}$) is chosen, and the $1$-$\sigma$ error is adopted from the FWHM of this probability distribution.

\subsection{Choice of indices}
Different from \citet{Thomas10a} we discard badly calibrated indices from the start. In Paper~I we find that the set of indices that appears to be best calibrated and most suited for the present aims are the Balmer line indices \HdA, \HgA, and \HgF,  the metallic indices \CNone, \CNtwo, Ca4227, G4300, \Ctwo, \Mgone, \Mgtwo, \Mgb, and the Fe indices Fe4383, Fe4531, Fe5270, Fe5335, and Fe5406. Then, similar to the approach of \citet{Graves08} we use different sets of indices for different elements.

\begin{figure*}
\includegraphics[width=0.45\textwidth]{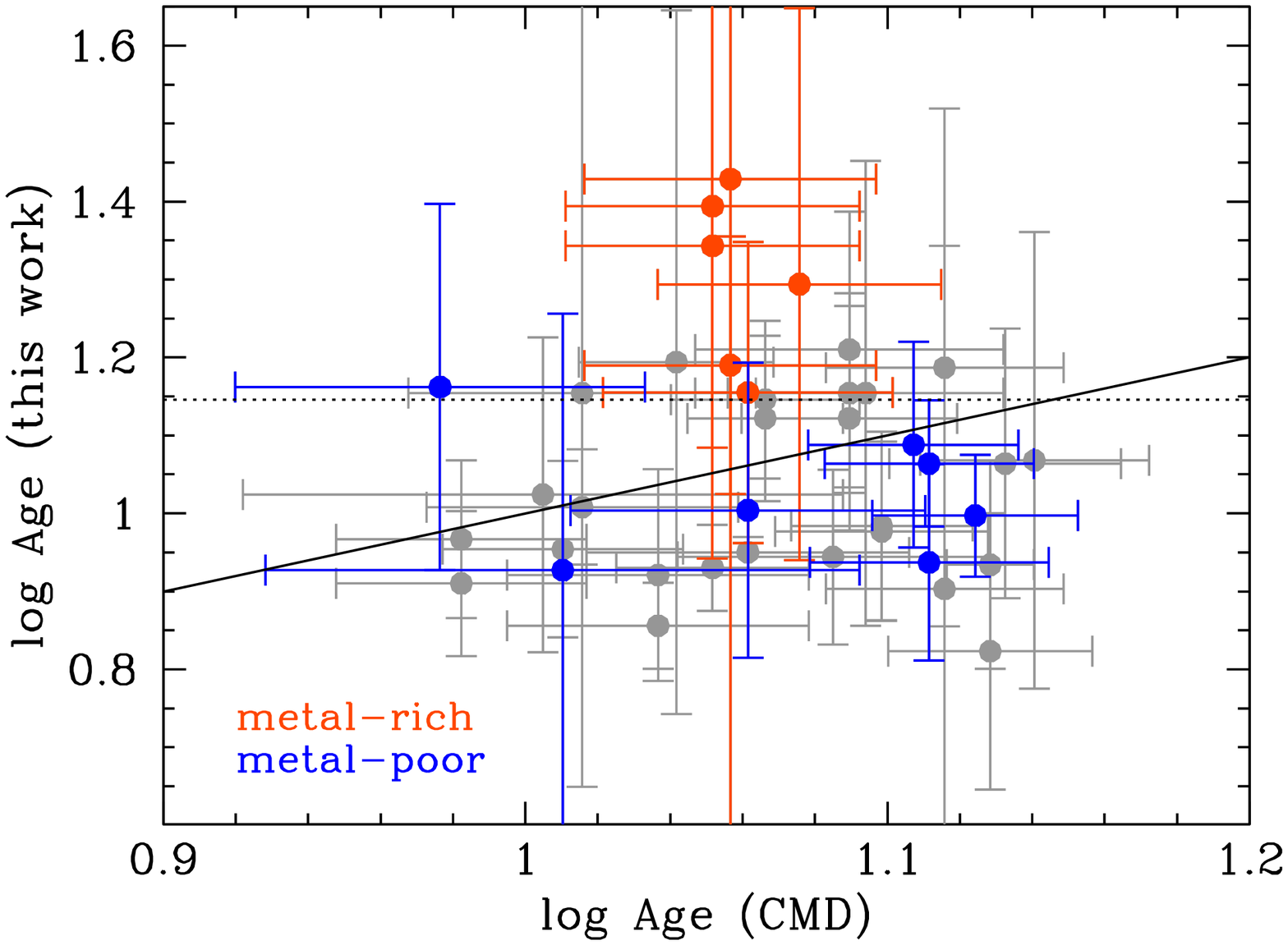}\hspace{0.05\textwidth}
\includegraphics[width=0.45\textwidth]{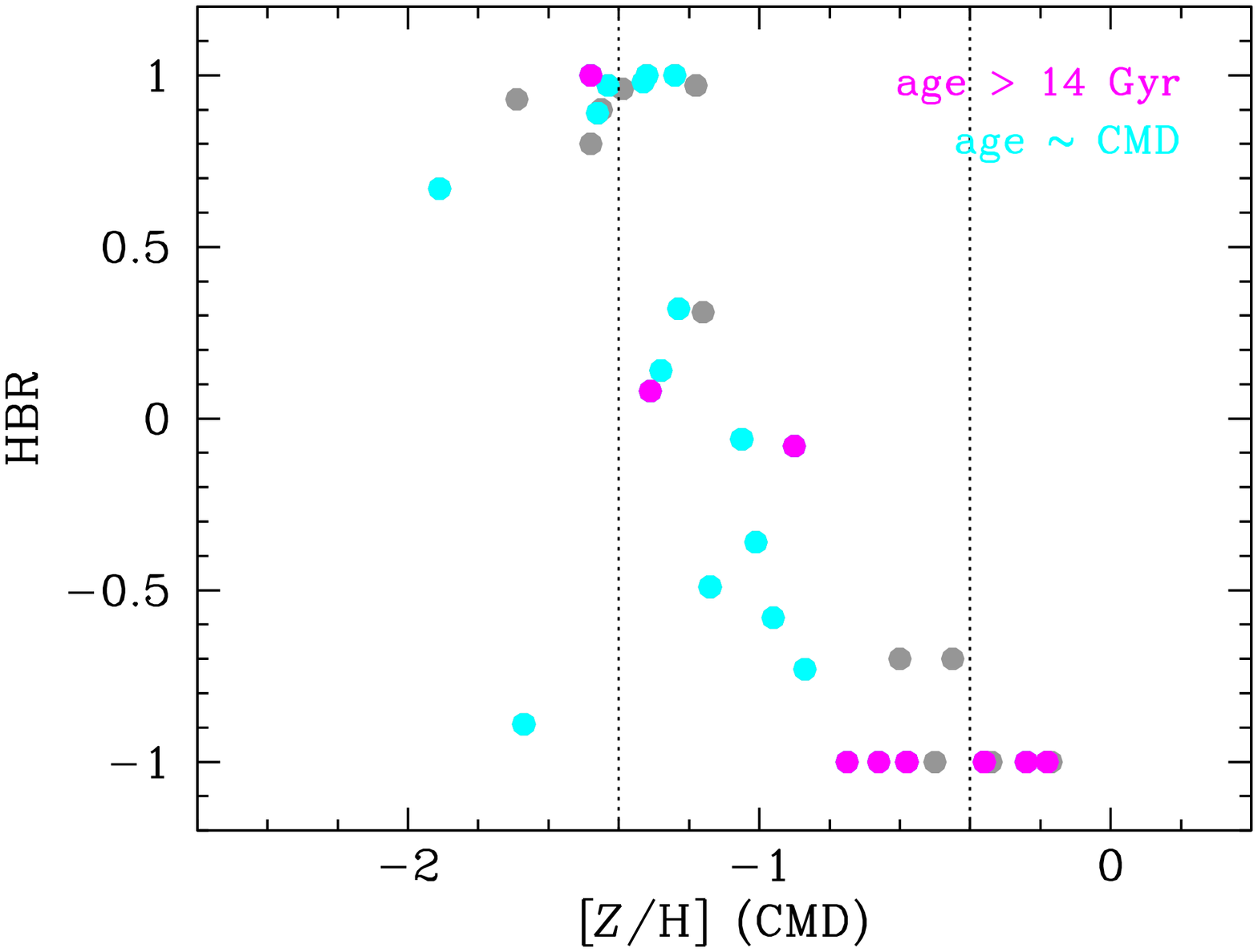}
\caption{Ages of galactic globular clusters derived from integrated light spectroscopy in comparison with literature data. Globular cluster spectra are taken from \citet{Puzia02} and \citet{Schiavon05}. Literature ages from colour-magnitude isochrone fitting are adopted from \citet{Marin09}. {\em Left-hand panel:} Grey symbols are the full sample, orange and blue symbols are metal-rich ($[\ZH]>-0.8\;$dex) and metal-poor ($[\ZH]<-1.55\;$dex) sub-samples, respectively. The dotted line marks the age of the universe as derived in \citet{Komatsu10}. {\em Right-hand panel}: Metallicity versus horizontal branch morphology (horizontal branch ratio HBR adopted from \citet{Harris96}). The literature metallicities on the \citet{ZW84} scale are taken from the compilation by \citet{Harris96}. The magenta symbols are those clusters for which we obtain ages larger than $14\;$Gyr, while the cyan symbols are clusters for which our ages agree with the CMD ages within $0.1\;$dex. The dotted lines are the metallicity limits from the left-hand panel. Literature ages are generally well reproduced. We tend to over-estimate ages for the most metal-rich globular clusters. Horizontal branch morphology only plays a minor role.}
\label{fig:agemet}
\end{figure*}

\subsection{Derivation of individual element abundances}
We define a base set of indices including \Mgb, the Balmer index \HdA, and the iron indices Fe4383, Fe5270, Fe5335, Fe5406. First we determine the traditional light-averaged stellar population parameters age, total metallicity, and \aFe\ ratio from this base set of indices. Only indices that are sensitive to these three parameters are included in the base set. In the subsequent steps we {\em add in turn} particular sets of indices that are sensitive to the element the abundance of which we want to determine. In each step we re-run the $\chi^{2}$ fitting code with a new set of models to derive the abundance of this element.
This new set of models is a perturbation to the solution found for the base set. It is constructed by keeping the stellar population parameters age, metallicity, and \aFe\ fixed and by modifying the element abundance of the element under consideration by $\pm 1\;$dex in steps of $0.05\;$dex around the base value.

A new best fit model is obtained from the resulting $\chi^{2}$ distribution. Then we move on to the next element. At the end of the sequence we re-determine the overall $\chi^{2}$ using all indices together and re-derive the base parameters age, metallicity, and \aFe\ for the new set of element ratios. Then we go back to the second step and use these new base parameters to derive individual element abundances. This outer loop is iterated until the final $\chi^{2}$ stops improving by more than 1 per cent. The method converges relatively fast and we generally require five steps at most to fulfil this criterion.

In more detail, the sequence of elements is as follows. The first element in the loop is carbon, for which we use the indices \CNone, \CNtwo, Ca4227, \HgA, \HgF, G4300, \Ctwo, \Mgone, and \Mgtwo\ on top of the base set. Next we drop these C-sensitive indices and proceed deriving N abundance, for which we use the N-sensitive indices \CNone, \CNtwo, and Ca4227. Then we move on to \Mgone\ and \Mgtwo\ for Mg abundance, Ca4227 for Ca, and finally Fe4531 for the element Ti. O abundance is indirectly inferred from the \aFe\ ratio by assuming that

\begin{equation}
[\OFe]\equiv [\aFe]\ .
\label{eqn:ofe}
\end{equation}

The typical errors for the parameters are $0.165\;$dex for log Age, $0.21\;$dex for [\ZH], $0.08\;$dex for [\OFe], and about $0.15\;$dex for the other element ratios. It should be emphasised again that these are very conservative error estimates.

\section{Results}
\label{sec:results}
From the $\chi^{2}$ fitting as described in Section~\ref{sec:model} we obtain age, total metallicity [\ZH], iron abundance [\FeH], the [\aFe] ratio, and the individual element abundance ratios [C/Fe], [N/Fe], [O/Fe], [Mg/Fe], [Ca/Fe], and [Ti/Fe] for a total of 52 globular clusters. We exclude the 47 Tucanae from our analysis, as age dating of this cluster from Balmer line indices is known to be problematic \citep{Schiavon02,Vazdekis01a}. In this section we present the results and compare with literature data obtained from colour-magnitude (CMD) fitting (age) and high-resolution spectroscopy of individual stars (metallicity and element abundance ratios). 

\subsection{Ages}
First we discuss the comparison of the derived ages with literature data. CMD ages are taken from \citet{Marin09} and \citet{DeAngeli05} where not available in \citet{Marin09}. The overlap of the two samples contains 39 clusters.

The age comparison is shown in the left-hand panel of Fig.~\ref{fig:agemet}, where we plot our derived ages as a function of the CMD ages. The grey symbols are the full sample, orange and blue are metal-rich and metal-poor sub-samples, respectively. The age derivation through the Lick indices works reasonably well. Almost half of the sample our globular cluster ages agree with \citet{Marin09} within $0.1\;$dex (18 out of 39), and three quarters (29) of the clusters agree within the (conservative) measurement errors. Most importantly, 35 clusters out of 52 (two thirds) are younger than the universe as derived by \citet[][dotted line]{Komatsu10} from a combination of cosmic microwave background, supernova, and large-scale structure data, and the vast majority (45 out of 52) are consistent with age of the universe within $0.1\;$dex. All but one cluster ages are consistent with the age of the universe within our (conservative) measurement errors.

Generally, the ages from Lick indices agree well with the CMD ages, albeit with quite a large scatter. Note that \citet{Mendel07} derived systematically larger ages with the TMB/K model\footnote{Note that we have not explicitly derived globular cluster ages in \citet{Thomas03a}. with respect to CMD. The major difference is that \citet{Mendel07} adopted CMD ages from \citet{DeAngeli05} who derived systematically younger absolute ages than \citet{Marin09}. It should be kept in mind, however, that the derivation of {\em absolute} globular cluster ages through CMDs carries its own problems as pointed out in both \citet{DeAngeli05} and \citet{Marin09}.} The turnoff brightness, being the major indicator for the age of a stellar population, is highly sensitive to the distance of the object. As a consequence, only relative ages can be reliably measured \citep{Ortolani95,DeAngeli05,Marin09}.

Still, it is interesting to investigate the reason for the exceedingly large ages of some clusters. Fig.~\ref{fig:agemet} shows that the clusters for which we overestimate the ages with respect to the age of the universe tend to be metal-rich with $[\ZH]>-0.4\;$dex (orange symbols). Most of the clusters whose ages agree well with the CMD ages, instead, are metal-poor with $[\ZH]<-1.4\;$dex (blue symbols). To investigate this further, in the right-hand panel of Fig.~\ref{fig:agemet} we plot metallicity (on \citet{ZW84} scale adopted from \citet{Marin09}) versus horizontal branch morphology expressed as horizontal branch ratio HBR adopted from \citet{Harris96}. The magenta symbols are those clusters for which we obtain ages larger than $14\;$Gyr, while the cyan symbols are clusters for which our ages agree with the CMD ages within the measurement errors. It can be seen clearly that horizontal branch morphology only plays a minor role at a given metallicity. Metallicity is the main driver for the age discrepancy. To summarise, globular cluster ages derived from absorption-line indices tend to be overestimated for the most metal-rich clusters around slightly sub-solar metallicity. This is a more general manifestation of the \Hb\ anomaly of globular cluster data noted by \citet{Poole10}, which may well be a 'Balmer anomaly' rather than being restricted to \Hb. Note, however, that this pattern is more likely to be caused by problems in the globular cluster data than the models, as galaxy data appear to be well matched instead \citep[][Paper~I]{Kuntschner10}.

\subsection{Metallicity}
\begin{figure}
\begin{center}
\includegraphics[width=\linewidth]{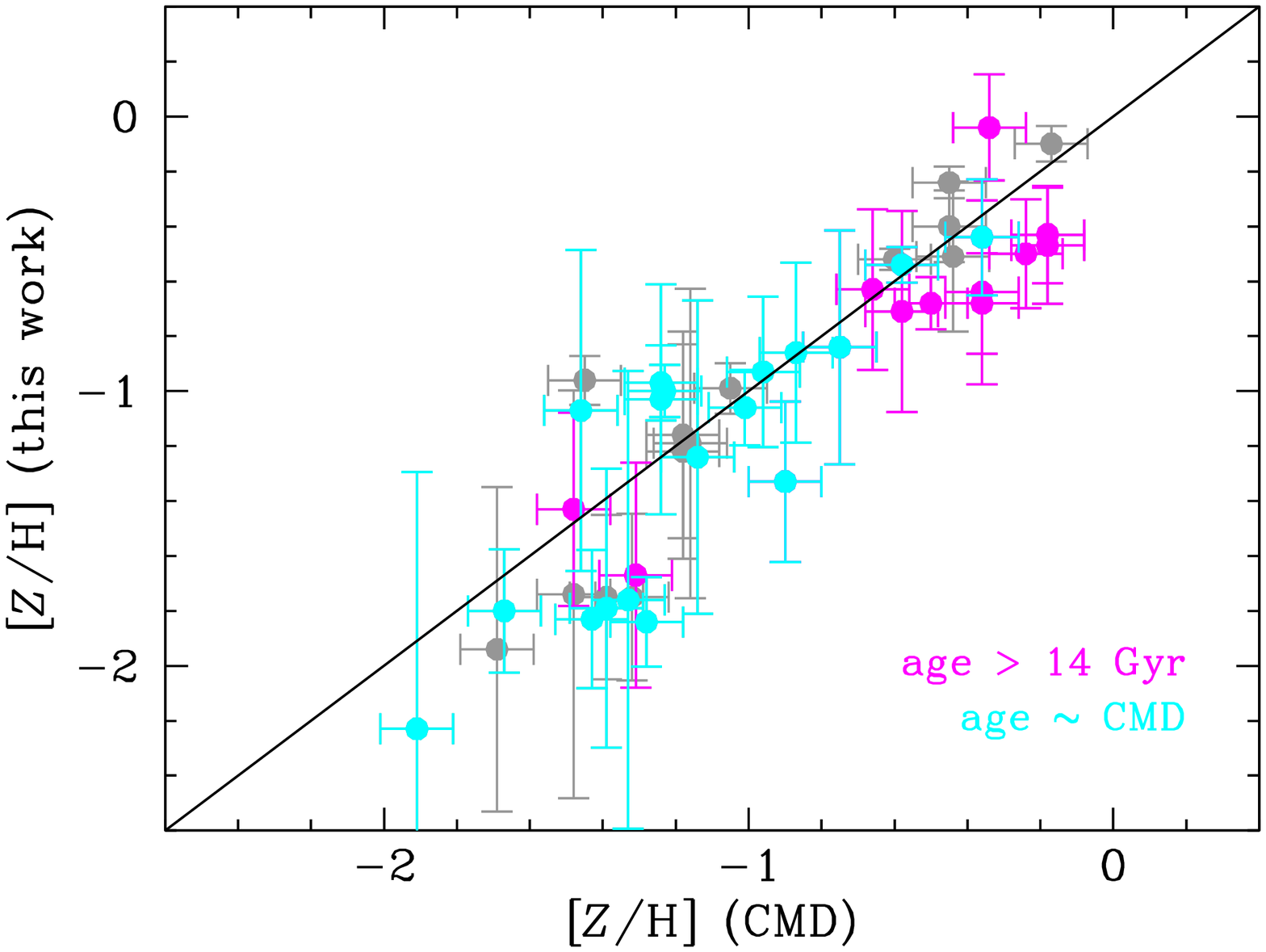}
\end{center}
\caption{Total metallicities [\ZH] of galactic globular clusters derived from integrated light spectroscopy in comparison with literature data. Globular cluster spectra are taken from \citet{Puzia02} and \citet{Schiavon05}. The literature metallicities on the \citet{ZW84} scale and corrected for $\alpha$-enhancement are taken from \citet{Marin09}. The magenta symbols are those clusters for which we obtain ages larger than $14\;$Gyr, while the cyan symbols are clusters for which our ages agree with the CMD ages within $0.1\;$dex. Literature metallicities are well reproduced. Metallicities are slightly underestimated for those clusters for which we overestimate the age.}
\label{fig:metallicity}
\end{figure}

For the comparison of metallicity we adopt literature values from \citet{Marin09} on the \citet{ZW84} scale. These represent total metallicity [\ZH] as \citet{Marin09} have corrected the iron measurement using the prescription of \citet{Salaris93}. We therefore confront these literature values with our measurements of total metallicity [\ZH] in Fig.~\ref{fig:metallicity}. The magenta symbols are those clusters for which we obtain ages larger than $14\;$Gyr, while the cyan symbols are clusters for which our ages agree with the CMD ages within $0.1\;$dex.

Metallicities agree very well, with a tendency of slightly lower metallicity estimates from the present work at low metallicities. It can further be seen from Fig.~\ref{fig:metallicity} that this match is particularly good for those clusters whose Lick index ages agree best with the CMD ages (cyan symbols). For the clusters with the oldest Lick index ages (magenta symbols), instead, we tend to underestimate metallicity by $\sim 0.2\;$dex This might in fact be an artefact produced by the age-metallicity degeneracy, i.e.\ we tend to underestimate the metallicity of those globular clusters for which we overestimate the age.

\begin{figure*}
\includegraphics[width=0.33\textwidth]{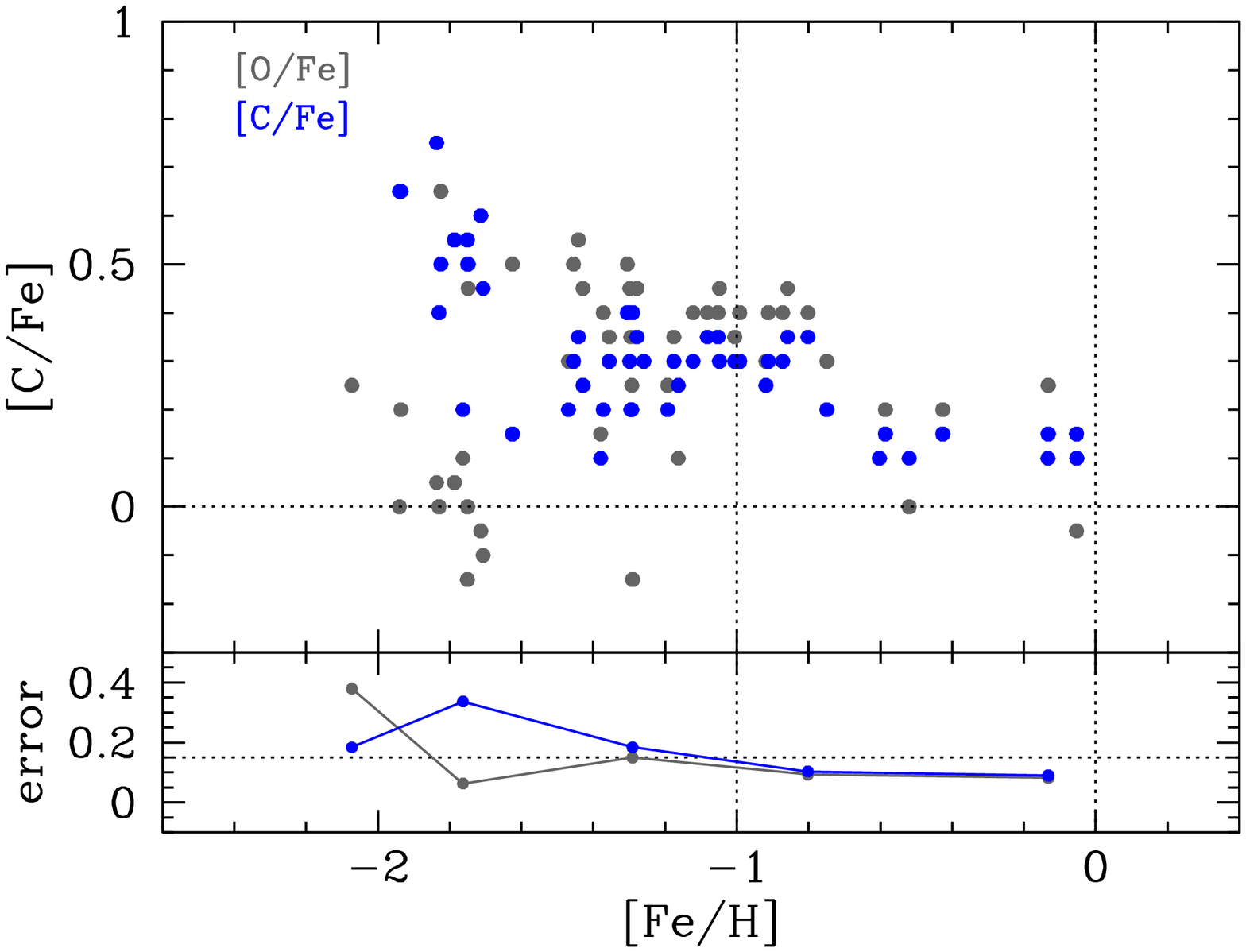}
\includegraphics[width=0.33\textwidth]{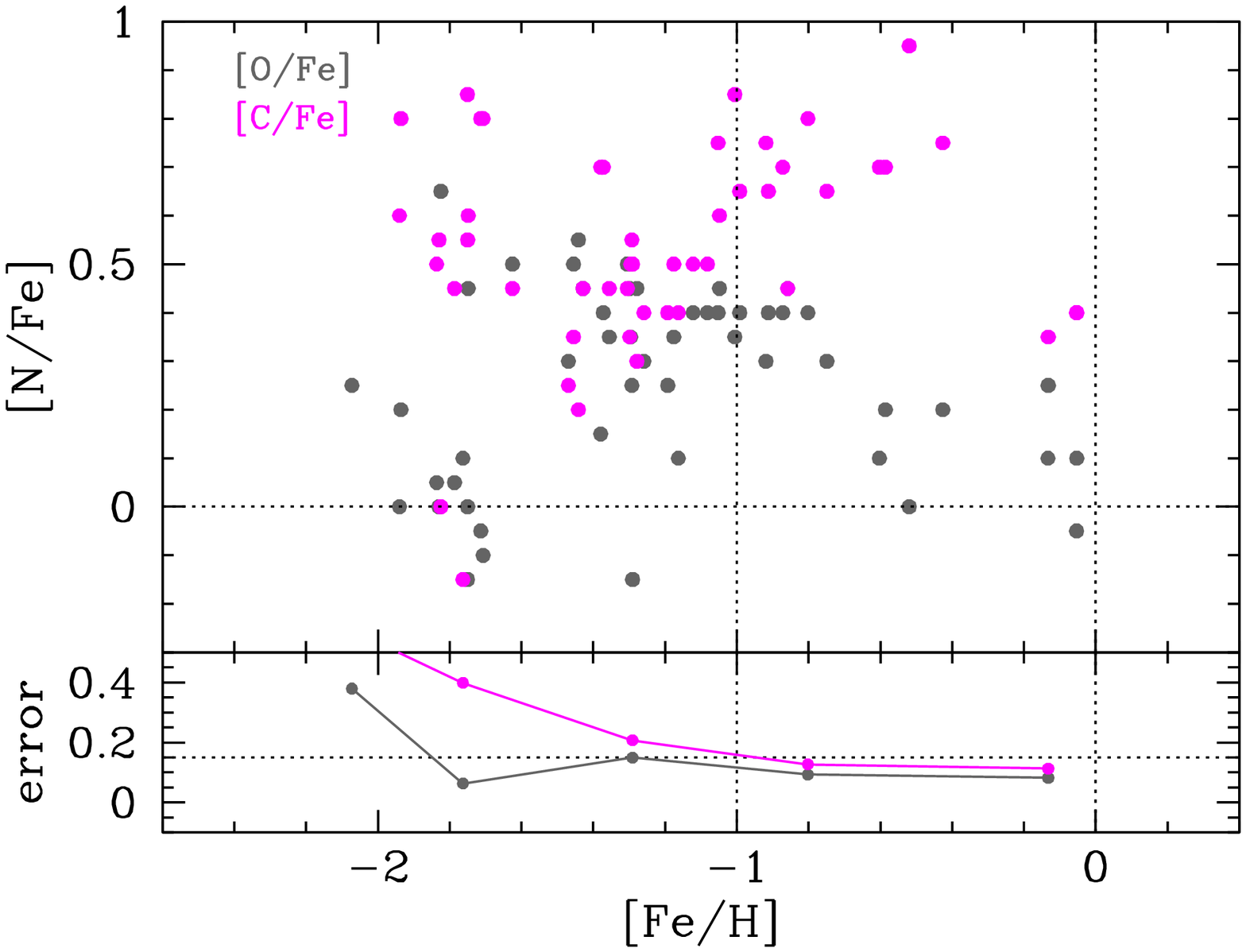}
\hspace{0.33\textwidth}\mbox{}
\includegraphics[width=0.33\textwidth]{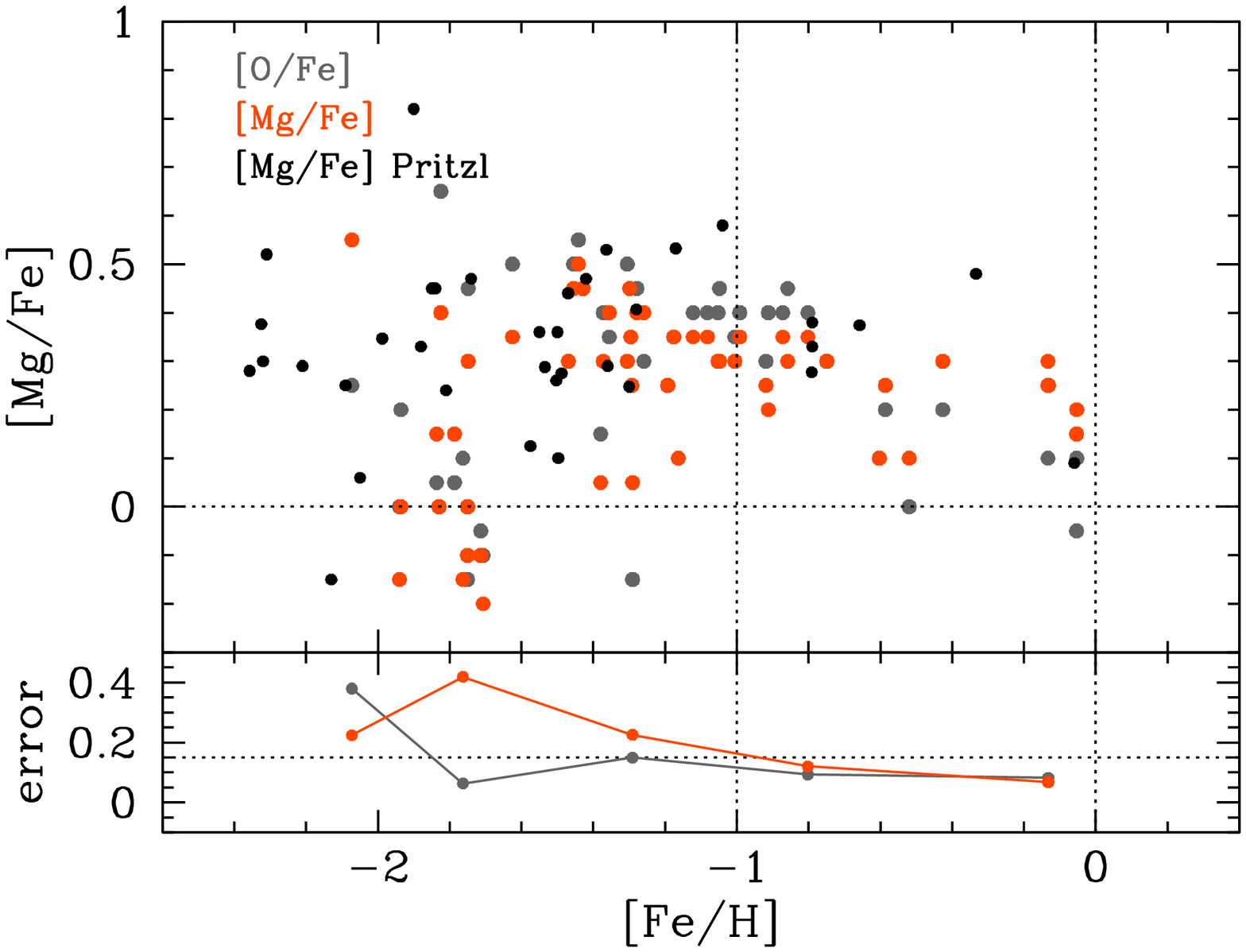}
\includegraphics[width=0.33\textwidth]{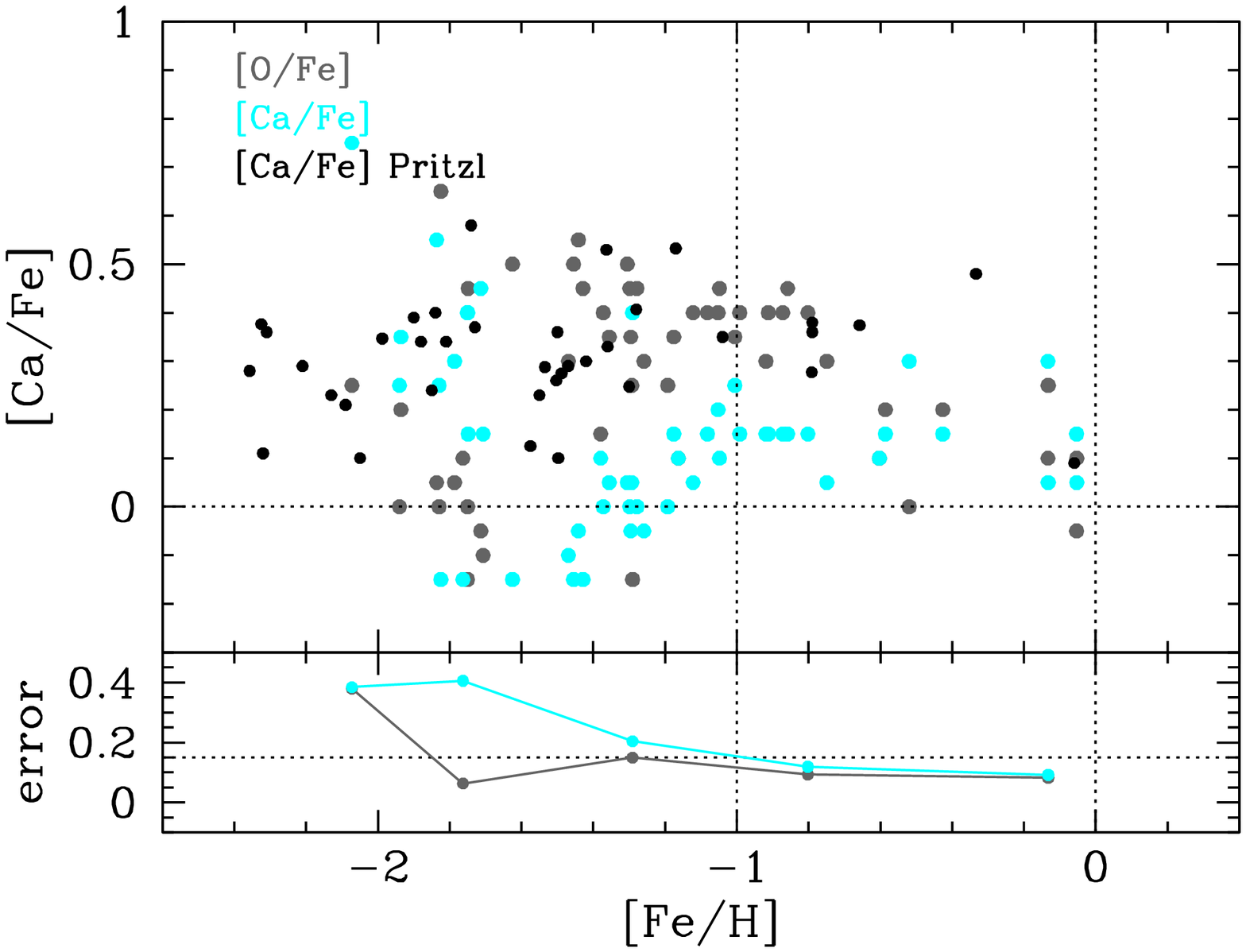}
\includegraphics[width=0.33\textwidth]{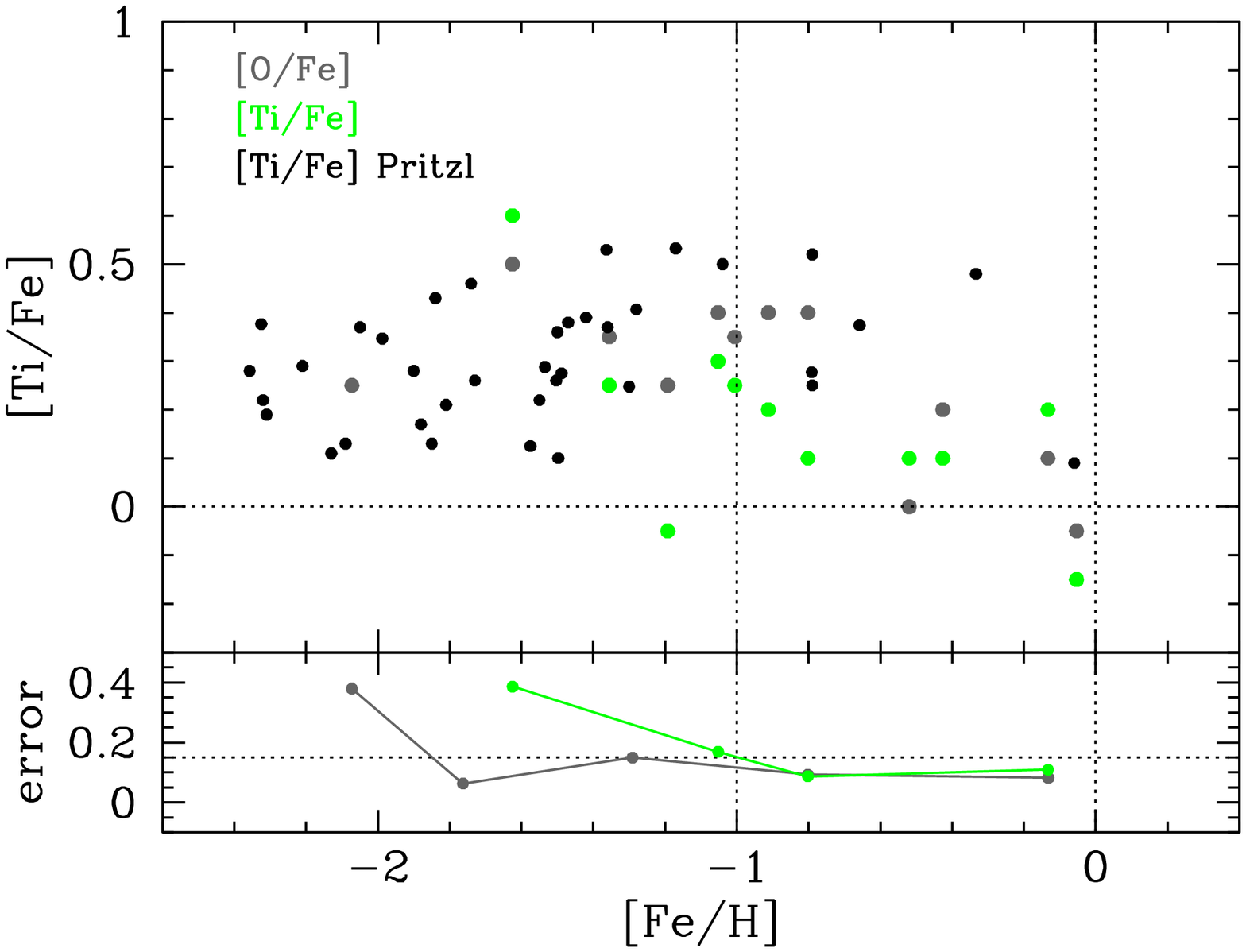}
\caption{Abundance ratios [C/Fe], [N/Fe], [Mg/Fe], [Ca/Fe], and [Ti/Fe] (coloured symbols) in comparison to the the [\OFe] ratio (grey symbols) as functions of iron abundance [\FeH] for galactic globular clusters. The globular cluster spectra are taken from \citet{Puzia02} and \citet{Schiavon05}. The black dots are the element ratios of globular cluster stars from \citet{Pritzl05}. The bottoms panels show the 1-$\sigma$ error on the element ratios with the dotted horizontal lines indicating an error of $0.15\;$dex. The typical abundance pattern of Milky Way field and globular cluster stars is well reproduced for [\OFe]. The other element ratios have too large errors at low metallicities below $[\FeH]\sim -1\;$dex, hence meaningful conclusions can only be drawn at $[\FeH]\ga -1\;$dex.}
\label{fig:xfefeh}
\end{figure*}

\subsection{Element abundance ratio pattern}
We now turn to discuss the individual abundances of the elements C, N, O, Mg, Ca, and Ti relative  to the abundance of Fe. Fig.~\ref{fig:xfefeh} presents the abundance ratios [\CFe], [\NFe], [\MgFe], [\CaFe], and [\TiFe] (coloured symbols) as functions of iron abundance [\FeH] in comparison to [\OFe] (grey symbols). the bottom panels indicate the typical measurement error as a function of iron abundance.

The element ratio [\OFe], being equivalent to [\aFe] (see equation~\ref{eqn:ofe}), carries the smallest measurement error. This element ratio is well determined at all metallicities. The expected pattern of super-solar [\aFe] with a slight decrease toward solar metallicity is reproduced. The individual element abundance ratios, instead, have significantly larger errors. In all cases the typical errors increase with decreasing metallicity, and exceed $\sim 0.1\;$dex at $[\FeH]<-1\;$dex. In fact, the abundance pattern loses structure at such low iron abundance. This ought to be expected as the sensitivity of the models to element ratio variations decreases dramatically with decreasing metallicity \citep[e.g.,][]{Thomas03a}. Moreover, model errors become comparable to model variations for only moderate abundance ratio changes (see Paper~I), which further hampers the analysis. Hence abundance ratios cannot be reliably determined. At $[\FeH]>-1\;$dex, instead, our results reveal interesting abundance trends.

\subsection{Comparison with stellar spectroscopy}
\begin{figure}
\includegraphics[width=\linewidth]{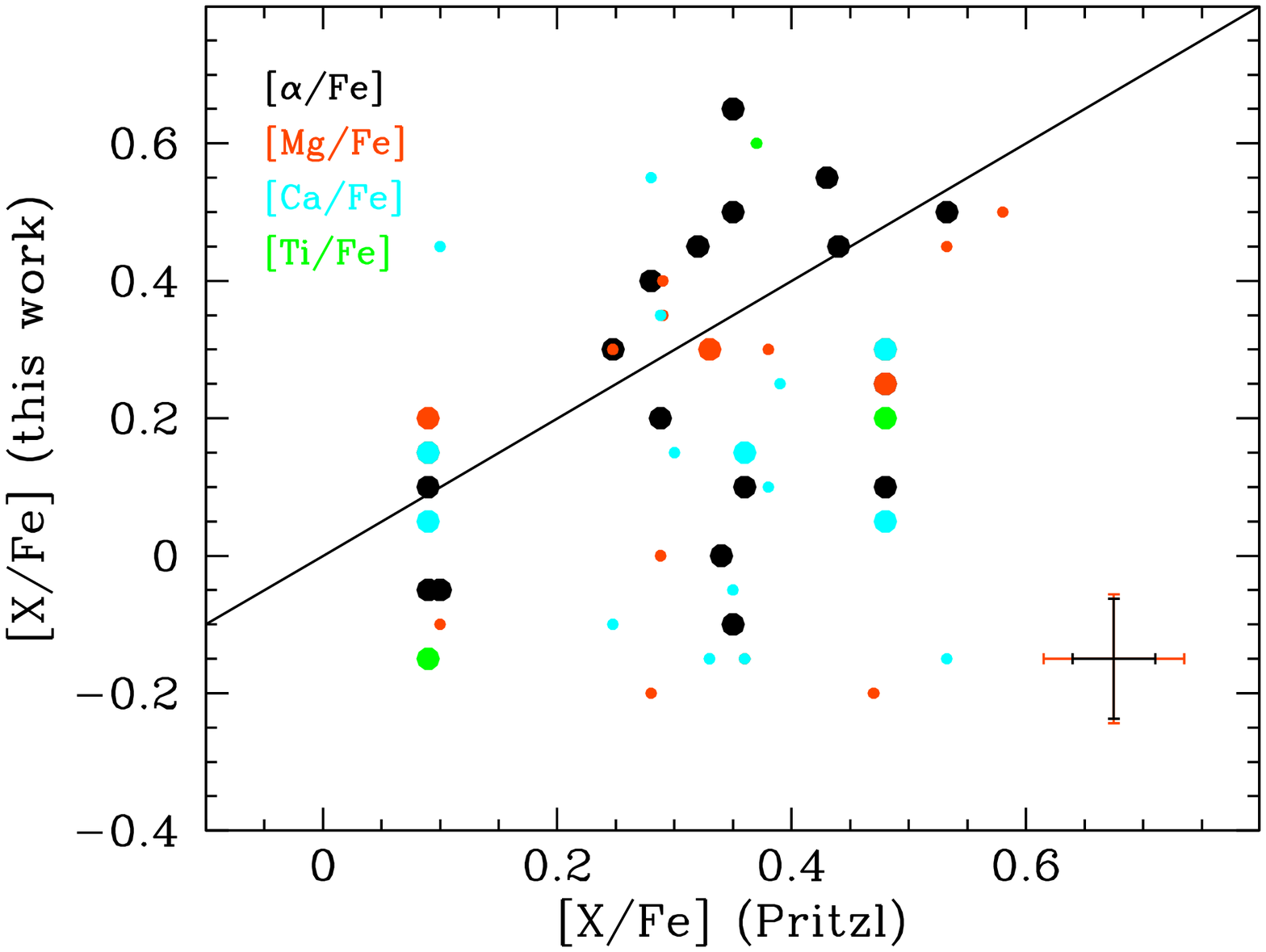}
\caption{Abundance ratios [Mg/Fe], [Ca/Fe], and [Ti/Fe] (coloured symbols) of galactic globular clusters measured in this work from integrated light spectroscopy in comparison with literature values from individual stellar spectroscopy by \citet{Pritzl05}. Black symbols show [\aFe] ratios. The error symbols indicate typical errors in both [\aFe] and [\MgFe]. The small symbols are globular clusters with $[\FeH]<-1\;$dex, for which element ratios from integratd light spectroscopy are unreliable (see Fug.~\ref{fig:xfefeh}).}
\label{fig:pritzl}
\end{figure}Before discussing these abundance patterns in detail, we present the direct comparison of our results with the measurements of \citet{Pritzl05}, who have derived element abundance ratios of a large sample of galactic globular clusters form individual stellar spectroscopy. \citet{Pritzl05} have observed between one and ten stars per cluster and derived the element ratios [\MgFe], [\CaFe], and [\TiFe]. The ratio [\aFe] is computed from the geometrical mean of these three measurements. We have computed the straight average when more than one star has been observed. In total, the \citet{Pritzl05} sample has 18 clusters in common with the present work.

In Fig.~\ref{fig:pritzl} we plot the abundance ratios [Mg/Fe], [Ca/Fe], and [Ti/Fe] (coloured symbols) as derived in the present work from integrated light spectroscopy as functions of the measurements from \citet{Pritzl05}. Black symbols show [\aFe] ratios. The error symbols indicate typical errors in both [\aFe] and [\MgFe]. The small coloured symbols are globular clusters with $[\FeH]<-1\;$dex, for which individual element ratios from integrated light spectroscopy are less reliable (see Fig.~\ref{fig:xfefeh}). There is a satisfactory agreement for [\aFe] at all metallicities, in agreement with the results of \citet{Mendel07} obtained with the TMB/K models. [\MgFe] ratios are still in reasonable agreement at $[\FeH]>-1\;$dex. There is a hint for systematically lower [\CaFe] and [\TiFe] ratios in our work, instead. We present a full discussion of this discrepancy in the following sections.

\subsection{Element abundance distributions}
\begin{figure*}
\includegraphics[width=0.33\textwidth]{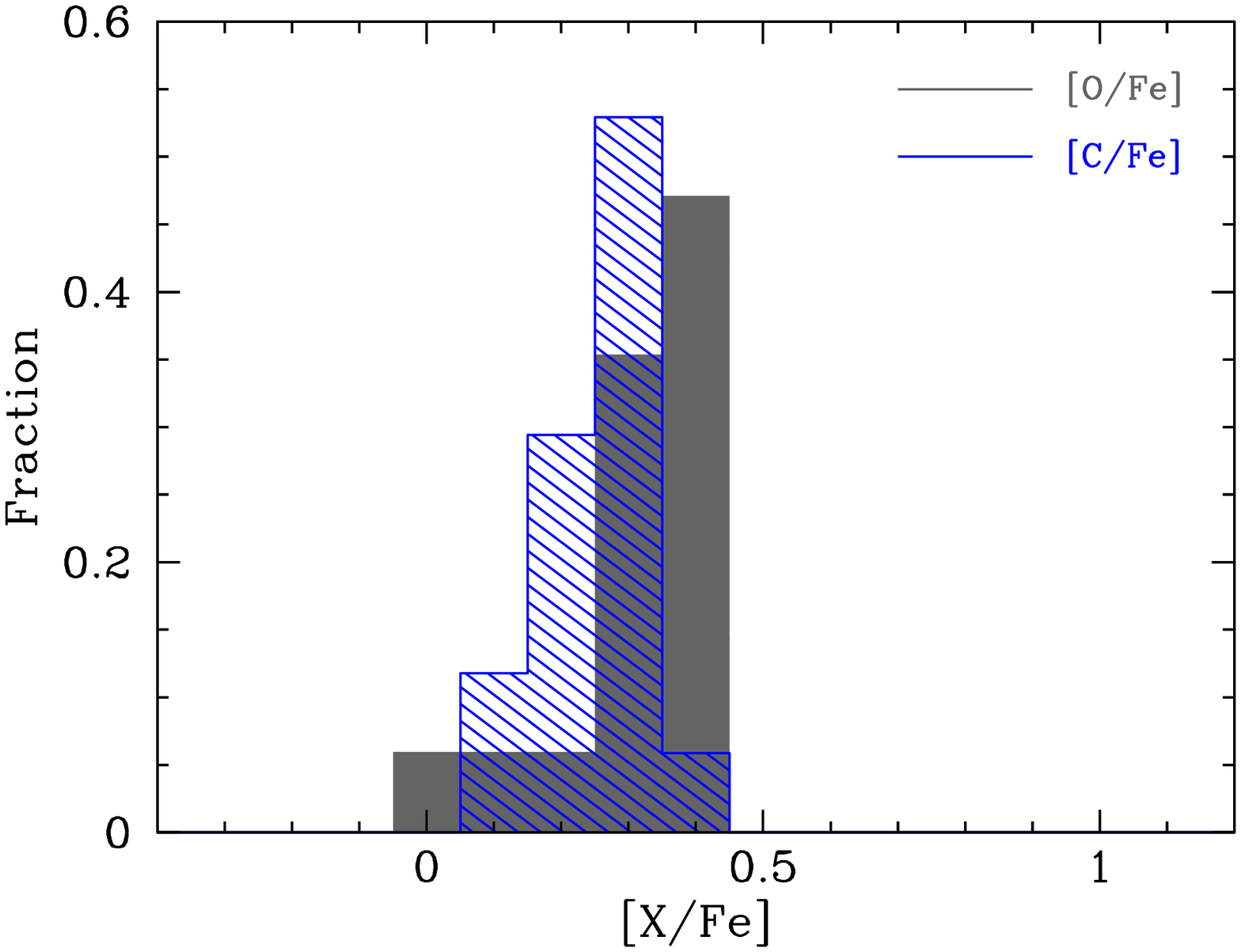}
\includegraphics[width=0.33\textwidth]{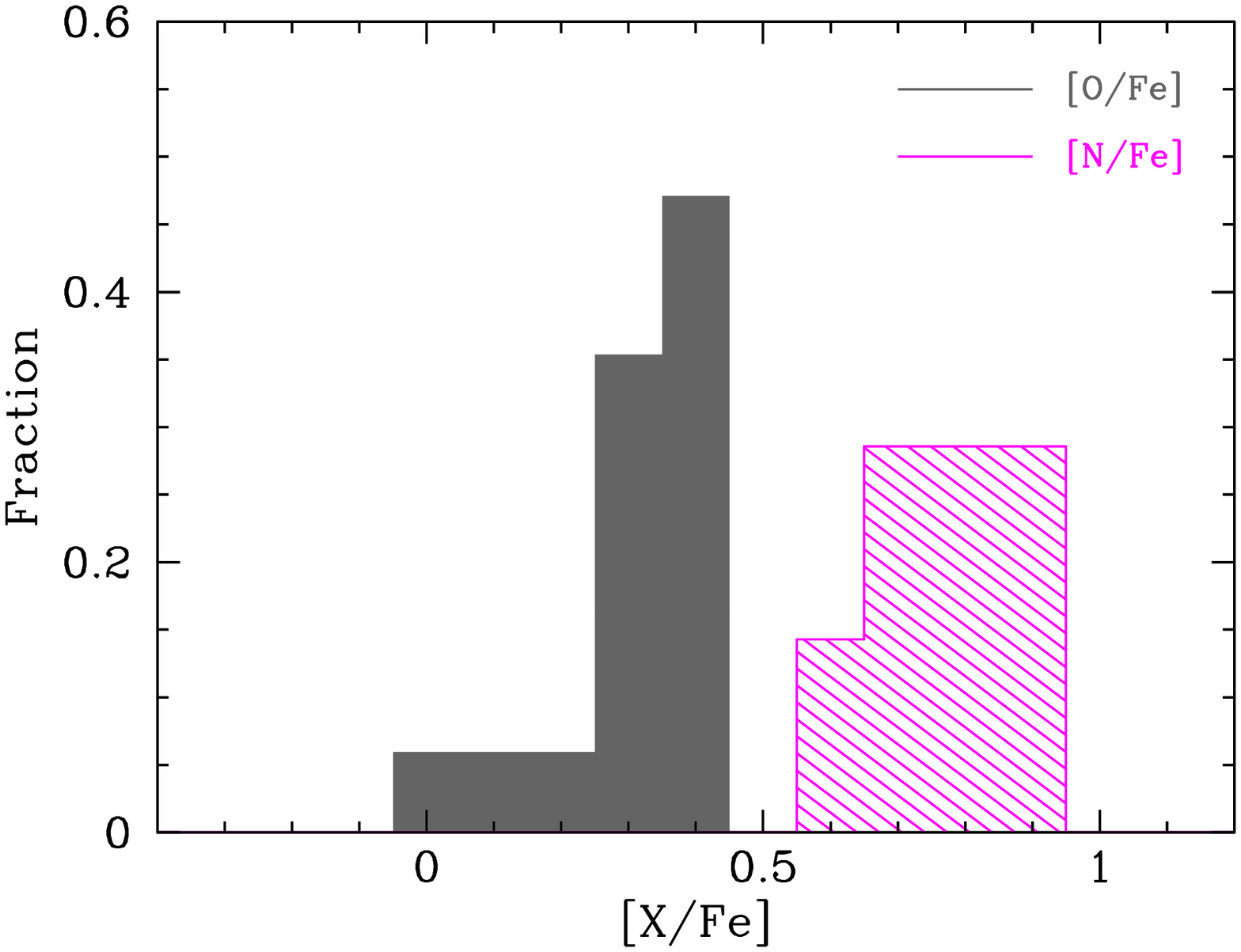}
\hspace{0.33\textwidth}\mbox{}
\includegraphics[width=0.33\textwidth]{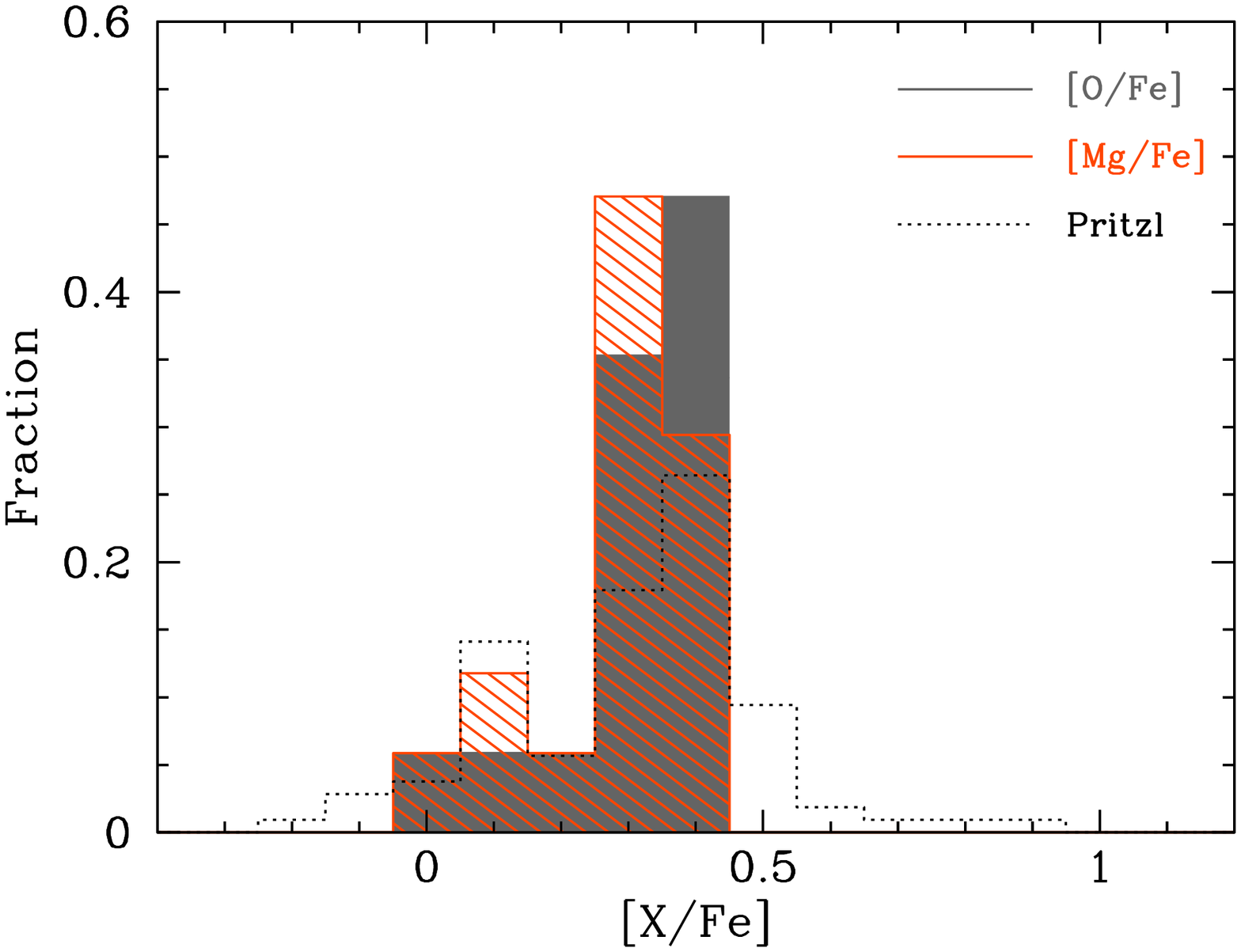}
\includegraphics[width=0.33\textwidth]{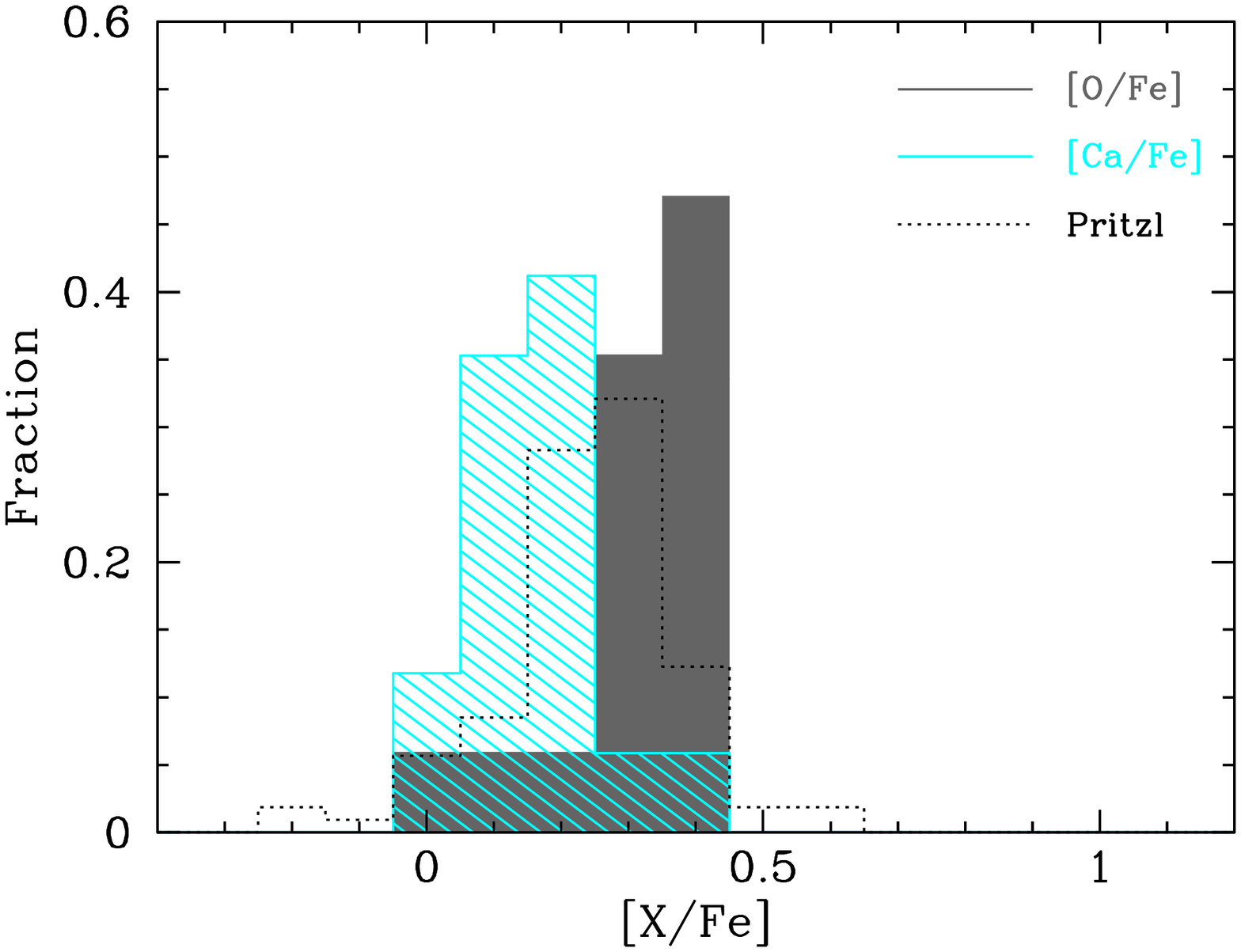}
\includegraphics[width=0.33\textwidth]{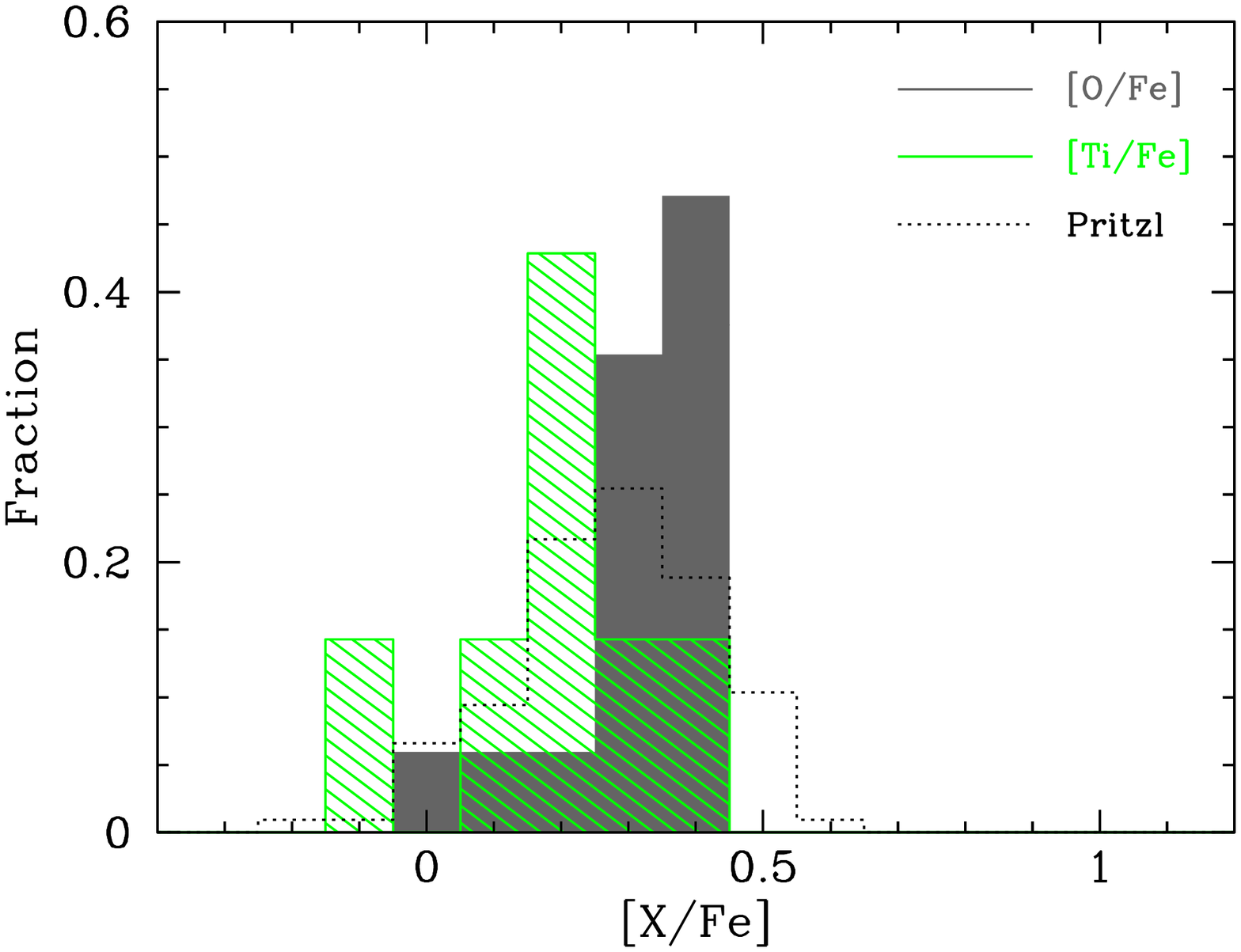}
\caption{Distributions of [C/Fe], [N/Fe], [Mg/Fe], [Ca/Fe], and [Ti/Fe] ratios (coloured histograms) in comparison to the distribution of the [\OFe] ratio (grey histogram) for galactic globular clusters. The globular cluster spectra are taken from \citet{Puzia02} and \citet{Schiavon05}. Only clusters with $[\FeH]> -1\;$dex are considered (15 objects out of 52) as element ratios cannot be reliably determined at lower metallicities (see text). The dotted black lines (for [Mg/Fe], [Ca/Fe], [Ti/Fe] only) are the distributions of the element ratios of globular cluster stars from \citet{Pritzl05}. We find a general trend such that the heavier of the light elements (Ca and Ti) are less enhanced than O and Mg. N is strongly enhanced in a sub-population of clusters accompanied by a slight depression of [C/Fe] with respect to O and Mg.}
\label{fig:elements}
\end{figure*}
In the following we compare the distributions of element ratios from the integrated and stellar spectroscopy. We only consider clusters with $[\FeH]> -1\;$dex in our analysis leaving us with a sample of 15 objects (out of 52). The reason is that element ratios cannot be reliably determined at lower metallicities, because the relative sensitivity of the model predictions to element ratio changes is too small (see Fig.~\ref{fig:xfefeh}). Note also that we only can consider the P02 sample for Ti (6 out of 12) as the Ti sensitive index Fe4531 cannot be measured for the S05 clusters.

\begin{figure*}
\includegraphics[width=0.8\textwidth]{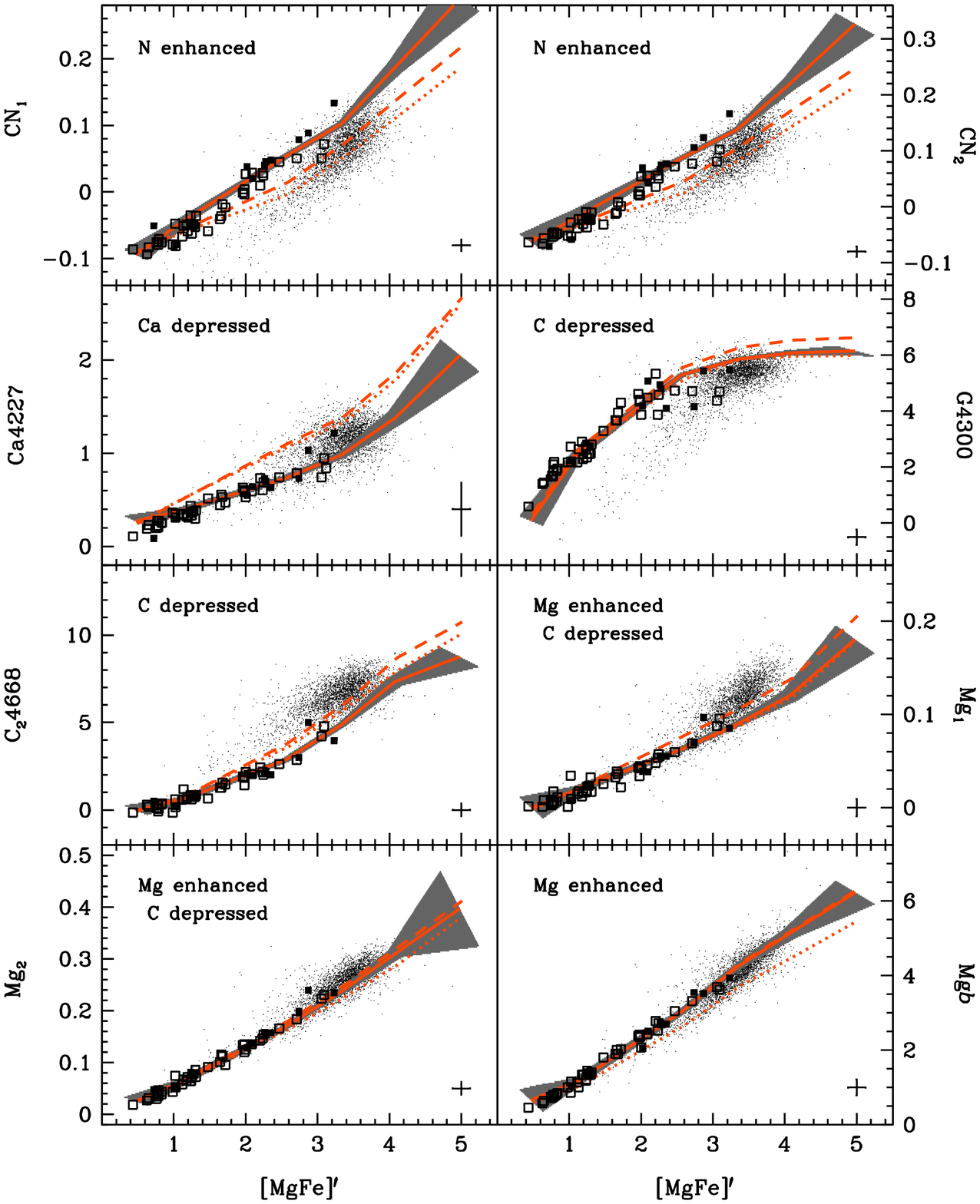}
\caption{Calibration of the line indices that are sensitive to light elements. Three models at Lick spectral resolution with an age of $13\;$ and the metallicities $[\ZH]=-2.25,\ -1.35,\ -0.33,\ 0.0,\ 0.35,\ 0.67\;$dex are shown. The solid lines are the final model for the average of the individual element abundance ratios derived through the $\chi^2$ fit. The dotted and dashed lines are the base models with $[\aFe]=0.0\;$dex and $[\aFe]=0.3\;$dex. The grey shaded area along the model indicates the 1-$\sigma$ error of the model prediction. Galactic globular clusters from \citet{Puzia02} and \citet{Schiavon05} are filled and open squares, respectively. The typical errors in the globular cluster index measurements are given the error symbol at the bottom of each panel. The small black dots are early-type galaxies from the MOSES catalogue \citep[MOrphologically Selected Early-type galaxies in SDSS][]{Schawinski07b,Thomas10a} drawn from the SDSS (Sloan Digital Sky Survey) data base \citep{York00} including only high signal-to-noise spectra with $S/N > 40$.}
\label{fig:enhancedel}
\end{figure*}
\begin{figure*}
\includegraphics[width=0.8\textwidth]{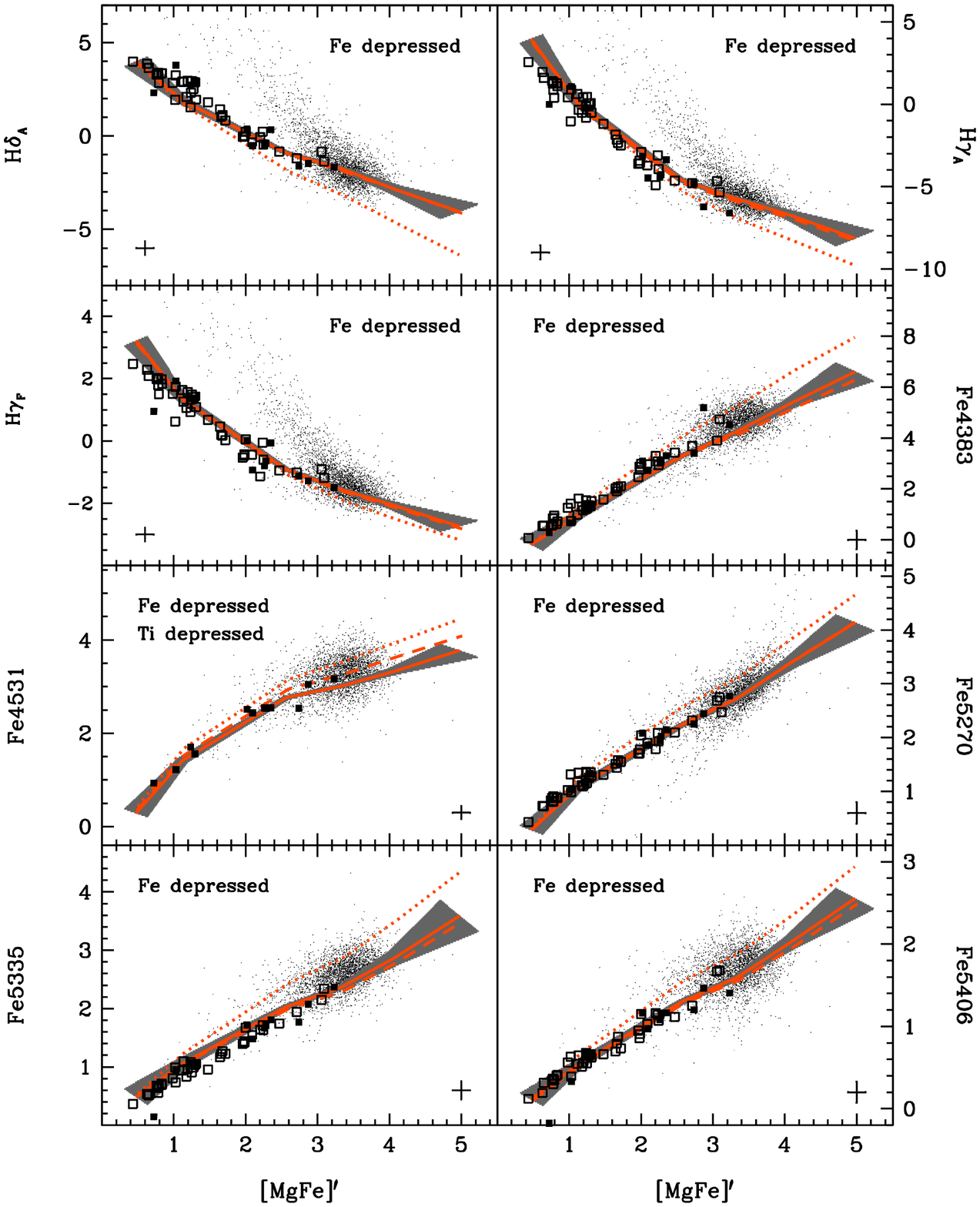}
\caption{Calibration of the Fe and Balmer line indices. Three models at Lick spectral resolution with an age of $13\;$ and the metallicities $[\ZH]=-2.25,\ -1.35,\ -0.33,\ 0.0,\ 0.35,\ 0.67\;$dex are shown. The solid lines are the final model for the average of the individual element abundance ratios derived through the $\chi^2$ fit. The dotted and dashed lines are the base models with $[\aFe]=0.0\;$dex and $[\aFe]=0.3\;$dex. The grey shaded area along the model indicates the 1-$\sigma$ error of the model prediction. Galactic globular clusters from \citet{Puzia02} and \citet{Schiavon05} are filled and open squares, respectively. Note that the indices Fe4531 and Fe5015 cannot be measured for the \citet{Schiavon05} sample. The typical errors in the globular cluster index measurements are given the error symbol at the bottom of each panel. The small black dots are early-type galaxies from the MOSES catalogue \citep[MOrphologically Selected Early-type galaxies in SDSS][]{Schawinski07b,Thomas10a} drawn from the SDSS (Sloan Digital Sky Survey) data base \citep{York00} including only high signal-to-noise spectra with $S/N > 40$.}
\label{fig:depressedel}
\end{figure*}

Fig.~\ref{fig:elements} shows the distributions of [C/Fe], [N/Fe], [Mg/Fe], [Ca/Fe], and [Ti/Fe] ratios (coloured histograms) in comparison with the distribution of the [O/Fe] ratio (grey histograms). The median values of these distribution are given in Table~\ref{tab:median}. The distribution of [\OFe] is reasonably tight with a median value of $0.24\;$dex as expected for Milky Way globular clusters. The other of the light $\alpha$ elements considered, [Mg/Fe], follows this distribution very closely with a very similar median of $0.25\;$dex. The other light elements, instead, deviate from this pattern. [C/Fe] and [Ca/Fe] ratios show similarly peaked distributions, but with different median values. The element ratio [C/Fe] has a median slightly lower by $\sim 0.04\;$dex, while [Ca/Fe] peaks at significantly lower values with a median of $0.15\;$dex. A Kolmogorov-Smirnov test confirms that the distributions in [\OFe] and [Ca/Fe] come from different underlying distributions at the $>6\sigma$ level. The distribution of [N/Fe] has a pronounced peak at a significantly larger value of $0.71\;$dex. The distribution of [Ti/Fe] is somewhat scattered. Still, the data show a clear trend toward lower [Ti/Fe] ratios with a median of $0.09\;$dex in line with the neighbouring $\alpha$ element Ca.

\begin{table}
\caption{Median values of element ratio distributions (in dex).}
\begin{tabular}{cccccc}
\hline
[O/Fe] & [C/Fe] & [N/Fe] & [Mg/Fe] & [Ca/Fe] & [Ti/Fe]\\
\hline
0.24 & 0.20 & 0.71 & 0.25 & 0.15 & 0.09\\
\hline
\end{tabular}
\label{tab:median}
\end{table}

The significant enhancement of nitrogen together with the slight depression of carbon relative to the other light elements is a well known abundance pattern in globular clusters observed in high-resolution spectroscopy studies of individual stars \citep[e.g.,][]{Norris84,Carretta05}. This chemical anomaly is commonly attributed to self-enrichment during the formation of the star cluster \citep{Ventura09}. Such N enhancement has been quantified in \citet{Thomas03a} for the first time for integrated light observations of globular clusters, while the accompanying depression of C found in the present work is new. The [C/Fe] and [N/Fe] ratios derived here appear to be well consistent with the measurements from \citet{Carretta05}.

The next heavier of the light elements, Mg, follows the distribution of [O/Fe] closely. This ought to be expected as these elements are close in atomic number and created in very similar processes during supernova nucleosynthesis \citep{WW95,TNH96}. However, the heavier $\alpha$ elements Ca and Ti deviate from this pattern. The typical [Ca/Fe] ratio is significantly lower than the typical [O/Fe] and [Mg/Fe] ratios. The [\TiFe] ratio is less well determined, but the results suggest that this element continues this trend with even lower [\TiFe] ratios.

\subsection{New model fits}
The adjustment of individual element abundances helps to improve the fits to a number of indices. In Figs.~\ref{fig:enhancedel} and \ref{fig:depressedel} we revisit the calibration figure from Paper~I for the model after the full chemical analysis. We only show those indices that have been used in the analysis, plotting index strengths as functions of \MgFep. Three models at Lick spectral resolution with an age of $13\;$Gyr are shown for the metallicities $[\ZH]=-2.25,\ -1.35,\ -0.33,\ 0.0,\ 0.35,\ 0.67\;$dex. Metallicity increases from left to right. The solid lines are the final model for the average of the individual element abundance ratios derived through the $\chi^2$ fit. The dotted and dashed lines are the base models with $\aFe=0.0\;$dex and $\aFe=0.3\;$dex for comparison. The grey shaded area along the model indicates the 1-$\sigma$ error of the model prediction. Galactic globular clusters from P02 and S05 are filled and open squares, respectively. The typical errors in the globular cluster index measurements are given the error symbol at the bottom of each panel. The small black dots are early-type galaxies from the MOSES catalogue \citep[MOrphologically Selected Early-type galaxies in SDSS][]{Schawinski07b,Thomas10a} drawn from the SDSS (Sloan Digital Sky Survey) data base \citep{York00} including only high signal-to-noise spectra with $S/N > 40$.

\subsubsection{Light element indices}
Fig.~\ref{fig:enhancedel} shows the indices that are sensitive to light element abundances, namely \CNone, \CNtwo, Ca4227, G4300, \Ctwo, \Mgone, \Mgtwo, and \Mgb. It can be seen from the top panels that the strengths of the CN indices are clearly underestimated in both solar-scaled and \aFe\ enhanced base models (dotted and dashed lines). As already discussed in \citet{Thomas03a} a significant enhancement in N is required to explain the high index strengths. At the same time, the indices G4300, and \Ctwo, and \Mgone\  are slightly too strong in the base models, which leads to a slight reduction of C abundance in the final best-fitting model. 

Another striking element abundance pattern that can be inferred from Fig.~\ref{fig:enhancedel} directly is the abundance of Ca. The strength of the index Ca4227 is significantly over-predicted by the solar-scaled and \aFe\ enhanced base models. The model matches the globular cluster data very well, instead, when a depression of Ca abundance is included. Finally, the indices \Mgtwo\ and \Mgb\ (bottom panels) are well reproduced by the \aFe\ enhanced model (dashed line), and only a negligible adjustment of Mg abundance is required to optimise the fit.

\subsubsection{Iron and Balmer line indices}
Fig.~\ref{fig:depressedel} presents the Fe and Balmer line indices used in the fitting procedure. The solar-scaled model (dotted lines) generally over-predicts the index strengths of the Fe indices, which is remedied through a depression of Fe abundance in the \aFe\ enhanced model (dashed lines). The index strength of Fe4531 is slightly re-adjusted through a depression of Ti abundance. The signal is very weak, though, and the determination of Ti abundance in this work must in fact be considered tentative, in particular since only a handful of clusters from P02 are available for the Ti abundance measurement. In general, the full chemical model only leads to minor corrections of the Fe indices. The same is true for the Balmer line indices. Here the solar-scaled model under-predicts line strengths, which is remedied by the enhancement of the \aFe\ ratio. Again, other elements only have negligible impact on these indices.

\section{Discussion}
\label{sec:discussion}
We have derived, for the first time, detailed chemical element abundance patterns of galactic globular clusters from integrated light spectroscopy. The light elements O and Mg show the well-known enhancement with respect to Fe, hence $[\OFe]\sim [\MgFe]\sim 0.3\;$dex. For C, N, and the heavier $\alpha$ elements Ca and Ti, however, we detected interesting abundance anomalies. N is further enhanced to very high [\NFe] ratios, while C is slightly depressed. Ca exhibits significantly lower [\CaFe] ratios than O or Mg, a pattern that appears to be present also in [\TiFe]. These anomalies have interesting consequences for supernova nucleosynthesis and the chemical enrichment in the Milky Way.

First we confront these results with the element ratios of individual stars in globular clusters as measured by \citet{Pritzl05}. These are shown by the dotted lines in Fig.~\ref{fig:elements} for [Mg/Fe], [Ca/Fe], and [Ti/Fe]. The distributions of the [Mg/Fe] ratios agree very well. [Ca/Fe] and [Ti/Fe] ratios, instead, are somewhat higher in \citet{Pritzl05}. The latter do suggest slightly lower enhancement of these elements, but not as pronounced as found here. But our finding gets support from the study by \citet{Feltzing09} who analyse six horizontal branch stars in the metal-rich galactic globular cluster NGC 6352. As expected the cluster is enhanced in the $\alpha$-elements. But like in the present work \citet[][see their Fig.~7]{Feltzing09} find a sequence of decreasing element ratios relative to iron for increasing atomic numbers from Mg through Ca to Ti.

It should be expected that field stars show the same behaviour since globular cluster element abundances generally follow the ones of the field stars in the galactic halo and discs \citep{Pritzl05}. In fact the trend reported here starts to crystallise out now from recent high quality stellar spectroscopy of galactic field stars in bulge and disc. \citet{Bensby10} analyse bulge and thick/thin disc stars and find [Ca/Fe] and [Ti/Fe] ratios to be lower than [O/Fe] and [Mg/Fe] ratios for all three populations.

The implication is that some fraction of the abundance in the heavier $\alpha$ elements must come from Type~Ia supernova explosions, while the lighter elements O and Mg remain to be enriched exclusively by Type~II. The yields of the W7 model for Type~Ia supernova explosions do indeed predict the production of traces of the heaviest $\alpha$ elements. As a consequence, galactic chemical evolution models predict lower [Ca/Fe] ratios for halo stars \citep{Chiappini97}, which had not been confirmed from observational data so far. The results discussed here provide a new observational support for this pattern.

This is critical for the chemical enrichment histories of galaxies. It leads to the most natural explanation for the shallow slope of the [\CaFe]-galaxy mass relation of early-type galaxies \citep{Saglia02,Thomas03b,Cenarro03,Michielsen03}. In this scenario, Ca is underabundant relative to the lighter $\alpha$ elements in massive galaxies for the same reason as Fe is underabundant \citep[see discussion in][]{Thomas03b}. The short formation time-scales inhibit Type~Ia supernovae to play a role in the chemical enrichment history of the stellar populations in these galaxies, such that elements that are produced in Type~Ia supernova are depleted in the stars. In fact chemical evolution models of bulges and spheroids predict lower element abundances for the heavier $\alpha$ elements, Ca in particular \citep{Matteucci99}. This implies that also Ti would have to be underabundant in massive galaxies. We investigate this issue in a companion paper (Johansson et al, in preparation).

\section{Conclusions}
\label{sec:conclusions}
Modelling integrated light spectroscopy of unresolved stellar populations allows us to study the detailed element abundances in distant galaxies and globular clusters. In Paper~I we present new, flux-calibrated stellar population models of Lick absorption-line indices with variable element abundance ratios (TMJ models). The new model includes a large variety of individual element variations, which allows the derivation of  the abundances for the elements C, N, O, Mg, Ca, Ti, and Fe besides total metallicity and age. In the present paper we use this model to obtain estimates of these parameters and element abundance ratios from integrated light spectroscopy of galactic globular clusters.

The globular cluster data is taken from \citet{Puzia02} and \citet{Schiavon05}. We measure line strengths of all 25 Lick absorption-line indices for both samples directly on the globular cluster spectra. Both globular cluster samples are flux calibrated, so that no further offsets need to be applied for the comparison with the TMJ models.

We derive the element abundance ratios [\CFe], [\NFe], [\OFe], [\MgFe], [\CaFe], [\TiFe] through an iterative $\chi^{2}$ fitting technique. First we determine the traditional light-averaged stellar population parameters age, total metallicity, and \aFe\ ratio from the indices \Mgb, the Balmer index \HdA, and the iron indices Fe4383, Fe5270, Fe5335, and Fe5406. In the subsequent steps we {\em add in turn} particular sets of indices that are sensitive to the element the abundance of which we want to determine. The indices used are \CNone, \CNtwo, Ca4227, \HgA, \HgF, G4300, \Ctwo, \Mgone, and \Mgtwo\ for carbon, \CNone, \CNtwo, and Ca4227 for nitrogen,  \Mgone\ and \Mgtwo\ for magnesium, Ca4227 for calcium, and Fe4531 for titanium. The Ti sensitivity of Fe4531 is relatively weak, hence the abundance derivations for this element are only tentative. The [\OFe] ratio is indirectly inferred by assuming that $[\OFe]\equiv [\aFe]$. We show that the model fits to these indices in globular clusters improve considerably through this full chemical analysis.

The ages we derive agree well with the literature. In particular the ages derived here are all consistent with the age of the universe within the measurement errors. There is a considerable scatter in the ages, though, and we overestimate the ages preferentially for the metal-rich globular clusters, which appears to extend the previously reported \Hb\ anomaly of globular clusters to the other Balmer indices.
Our derived total metallicities agree generally very well with literature values on the \citet{ZW84} scale once corrected for $\alpha$-enhancement, in particular for those cluster where the ages agree with the CMD ages. We tend to slightly underestimate the metallicity for those clusters where we overestimate the age, in line with the age-metallicity degeneracy.

It turns out that the derivation of individual element abundance ratios is highly unreliable at $[\FeH]<-1\;$dex, while the [\aFe] ratio is robust at all metallicities. The discussion of individual element ratios focuses therefore on globular clusters with iron abundances $[\FeH]>-1\;$dex.
We find general enhancement of light and $\alpha$ elements as expected with significant variations for some elements. The elements O and Mg follow the same general enhancement with almost identical distributions of [O/Fe] and [Mg/Fe]. We find slightly lower [C/Fe] and very high [N/Fe] ratios, instead. Hence N is significantly enhanced and C slightly depressed in globular clusters with respect to the other light elements. This chemical anomaly commonly attributed to self-enrichment is well known in globular clusters from individual stellar spectroscopy, and it is the first time that this pattern is derived also from the integrated light.

The $\alpha$ elements follow a pattern such that the elements with higher atomic number, namely Ca and Ti, are less enhanced. More specifically, [Ca/Fe] ratios are lower than [O/Fe] and [Mg/Fe] by about $0.2\;$dex. Ti continues this trend. We compare this result with recent determinations of element abundances in globular cluster and field stars of the Milky Way. We come to the conclusion that this pattern is now universally found. It suggests that Type~Ia supernovae contribute significantly to the enrichment of the heavier $\alpha$ elements as predicted in supernova explosion calculations and galactic chemical evolution models. This explains the presence of a Ca under-abundance (close to solar [Ca/Fe] ratios) in massive early-type galaxies and predicts similarly low [Ti/Fe] ratios in populations with short formation time-scales.

Tables with the absorption line indices, stellar population parameters and chemical element ratios of the globular clusters analysed here are available at www.icg.port.ac.uk/$\sim$thomasd.

\section*{Acknowledgements}
We would like to thank Harald Kuntschner, Ricardo Schiavon, and Gustav Str\"omb\"ack for the many stimulating discussions and for providing us with their galaxy and model data. We are deeply in dept with Alvio Renzini who pointed us to a mistake with the adopted CMD ages in the previous version. We thank the anonymous referee for very valuable comments that helped to improve the manuscript. CM acknowldges support by the Marie Curie Excellence Team Grant MEXT-CT-2006-042754 of the Training and Mobility of Researchers programme financed by the European Community.

Funding for the SDSS and SDSS-II has been provided by the Alfred P. Sloan Foundation, the Participating Institutions, the National Science Foundation, the U.S. Department of Energy, the National Aeronautics and Space Administration, the Japanese Monbukagakusho, the Max Planck Society, and the Higher Education Funding Council for England. The SDSS Web Site is http://www.sdss.org/.

The SDSS is managed by the Astrophysical Research Consortium for the Participating Institutions. The Participating Institutions are the American Museum of Natural History, Astrophysical Institute Potsdam, University of Basel, University of Cambridge, Case Western Reserve University, University of Chicago, Drexel University, Fermilab, the Institute for Advanced Study, the Japan Participation Group, Johns Hopkins University, the Joint Institute for Nuclear Astrophysics, the Kavli Institute for Particle Astrophysics and Cosmology, the Korean Scientist Group, the Chinese Academy of Sciences (LAMOST), Los Alamos National Laboratory, the Max-Planck-Institute for Astronomy (MPIA), the Max-Planck-Institute for Astrophysics (MPA), New Mexico State University, Ohio State University, University of Pittsburgh, University of Portsmouth, Princeton University, the United States Naval Observatory, and the University of Washington.



\bsp
\label{lastpage}

\end{document}